\documentclass[usenatbib,twocolumn]{mn2e}
\usepackage{graphicx,verbatim,amsmath,amssymb,enumitem}
\usepackage[table]{xcolor} 
\begin{document}
\topmargin-1cm
\title{
Mean maps for cosmic web structures in cosmological initial conditions
}
\author[Aung and Cohn]{Han Aung${}^{1}$ and J.D.Cohn${}^2$\\
${}^1$ Department of Physics and Department of Astronomy,
  University of California, Berkeley, CA 94720,\\
${} \; $ and
  Department of Physics, Yale University, New Haven, CT, 06520 \\
${}^2$ Space Sciences Laboratory and Theoretical Astrophysics Center,
  University of California, Berkeley, CA 94720\\
}
\maketitle
\date{\today}
\begin{abstract}
  Halos, filaments, sheets and voids in the cosmic web can be defined
  in terms of the eigenvalues of the smoothed shear tensor and a
  threshold $\lambda_{\rm th}$.  Using analytic methods, we construct
  mean maps centered on these types of structures for
  Gaussian random fields corresponding to cosmological initial
  conditions.  Each map also requires a choice of shear at the origin;
  we consider three possibilities.  We
  find characteristic sizes, shapes and other properties of the
central objects in these mean maps and explore how these properties
change with varying the threshold and
  smoothing scale, i.e. varying the separation of the cosmic web
  into different kinds of components.  The mean maps become
  increasingly complex as the threshold $\lambda_{\rm th}$ decreases
  to zero.  We also describe scatter around these mean maps,
subtleties which can arise in their construction,
 and some comparisons between halos in the maps and collapsed
 halos at final times.
\end{abstract}

\section{Introduction}
The web of cosmological large scale structure is both strikingly
evident in observations and simulations and
 difficult to characterize.  Its existence has been known for
several decades \citep{ZelEinSha82, ShaZel83, Ein84, BonKofPog96},
and a huge number of different approaches have been used to study it since.
Recent overviews of many of these directions can be found in the
Proceedings of the \citet{Tallinn14}.  

These cosmological structures evolved via gravitational
instability from initial conditions which have been 
characterized by measurements of the cosmic microwave
background (e.g., \citet{Pla13}).
These early time initial conditions are well 
described by Gaussian random fields, which can also be classified
into cosmic web structures.  
An assortment of different approximate methods can be
used to estimate properties of the late time counterparts of the structures
found in the initial conditions.  Properties include
the fraction of mass which will collapse into a certain type of
cosmological structure, the number of such structures, etc.
Classic works include \citet{PS,BBKS}; some
textbook descriptions are found in
\citet{Pea98} and \citet{MovdbWhi10}.
For instance, peaks in the initial conditions can be associated
with collapsed halos at late times on average \citep{Kai84}, and
properties of the peaks can be connected to final time properties
of the halos (for instance by tracing back particles in simulations
to their early time configurations such as in,
e.g., \citet{DalWhiBonShi08,Rob09,LudPor11,EliLudPor12,DesTorShe13,BorLudPor14,LudBorPor14}). 
Halos are the most studied cosmological structures, in part due to their
close identification to galaxies \citep{Blu84}.

This note extends studies of cosmic structures within the specific
cosmic web classification considered in \citet{AloEarPea14}.  (Their detailed
particular definitions are in \S \ref{sec:background} below.)  In that work, they
compared analytic predictions for initial random Gaussian fields
(equivalently, their values linearly extrapolated to final time) and
measurements of fully evolved dark matter simulations at late times
(redshift $z=0.1$).  For some values of their parameters, including
the primary one which we focus upon in this work, the volume fractions
of the structures for Gaussian random fields and structures for fully
evolved dark matter simulations counterparts at late times were
comparable.  The analytic mass fractions followed the same trends as
found in the final time simulations, but were somewhat too high for
sheets and too low for filaments.  (Halos, which have the largest
overdensities and thus might be expected to have the largest
differences in the two methods, contribute a very small fraction to
both the mass and volume fractions in this structure classification.)

This analytic classification of cosmological Gaussian initial conditions 
has further consequences, as the power spectrum also determines 
mean spatial properties around a given central point.
We analyze these spatial properties in this note.
We find mean properties of regions around different choices of origin,
classified
as halos, filaments, or sheets, regions we call mean halos, mean filaments and
mean sheets.   (The boundary around a ``mean void'' is
more challenging to define as for many parameter choices the mean
background is also a void, so that a boundary is not evident).  
The central shear values of these mean objects are an additional
parameter which must be chosen.  We  explore
three natural choices for these mean object central shear values, and
measure sizes, shapes, shear profiles, and mean linearly extrapolated
overdensities of the resulting mean halos, filaments and sheets.  We
also discuss scatter around these mean configurations, subtleties in
their construction, and some possible relations between mean halos and final time
collapsed halos.  
We consider structures in dark matter only.

Background is given in \S \ref{sec:background}:
the particular shear-based web classification we use, 
analytic estimates for average shear in different objects,  
the construction of mean maps and the measurement of some of their properties.
In \S\ref{sec:examp} we consider three choices of
central shear, showing
mean map examples for different objects and 
trends in their properties when changing the parameters defining the cosmic web.
We also discuss scatter around the mean.  
In \S\ref{sec:collapse} we turn to possible
connections of the mean halos to collapsed halos at late times.
In \S\ref{sec:conclude} we summarize and
conclude.
The three appendices give details of the calculations (shear,
Zel'dovich displacement and scatter), subtleties which arise
in some interpretations of one of our construction methods
and properties of mean maps corresponding to slightly different constraints
at the origin used by \citet{PapShe13}.   The power spectrum used is
for a $\Lambda$CDM universe with $(\Omega_m,\Omega_bh^2,h,\sigma_8,n) =
(0.31,0.022,0.68,0.83,0.96)$ \citep{Pla13}, calculated with the
\citet{EisHu99} transfer function.

\section{Definitions and methods}
\label{sec:background}

The cosmic web can be separated into halos (sometimes called 
nodes), filaments, sheets and
voids.  We will use the term halo and node interchangeably, as both
terms are commonly used for these structures.  There are many different web classifications
based upon density, morphology, shear and beyond, some of which
concentrate on just finding one particular type of
structure (e.g. halos or voids) and others which classify an entire map.  
We use a classification of the latter sort, 
assigning a type of cosmic structure to every region based upon its shear.

\subsection{Shear}
We begin with details of our shear definition and classification.  We follow 
\citet{DesSmi08,PapShe13} for definitions and for much of the
analysis methodology below.
The shear is defined on the initial Gaussian perturbations as
\begin{equation}
\xi_{ij} ({\bf r})= 
\frac{1}{\sigma}\partial_i \partial_j \Phi({\bf r}) \;. 
\label{eq:sheardef}
\end{equation}
Our potential $\Phi ({\bf r})$
has the same spatial dependence as the Newtonian potential $\phi$ at early
times but
is time independent: $\Phi ({\bf r}) = \phi({\bf r},a)/4 \pi G a^2 {\rho}_b D(a)
$.  Here $G$ is Newton's constant, $\rho_b$ is the mean
background density of the universe, and $D(a)$ is the linear growth factor
describing how linear structure grows as the universe expands with
scale factor $a$.
The derivatives $\partial_i$ are with respect to the comoving
coordinates ${\bf r}$.
The potential $\Phi({\bf r})$ is related to the initial density
perturbations evolved linearly to the present time,  $\delta({\bf r})$,  via the Poisson equation:
\begin{equation}
\sum_i \partial_i^2 \Phi ({\bf r}) = \delta({\bf r}) \; \;  .
\label{eq:delphi}
\end{equation}
At times earlier than the present time the linearly evolved
overdensity obeys
\begin{equation}
\delta_L({\bf r},a) = D(a) \delta({\bf r}) \; .
\end{equation}
The linearly evolved overdensity at the present time, 
$\delta ({\bf r})$ at position ${\bf r}$ is defined in
terms of (the linearly evolved to the present time) density via $\rho
({\bf r})/\bar{\rho} -1$.

The factor $\sigma$ is
the square root of the integral of the 
power spectrum $P(k)$ associated with $\delta({\bf r})$, smoothed on scale $R_s$
\begin{equation}
\sigma^2 = \frac{1}{2 \pi ^2} \int dk k^{2} P(k) W_{R_s}^2(k) \; ,
\label{eq:sigdef}
\end{equation}
i.e., the root mean square fluctuations of the
density contrast $\delta({\bf r})$. 
We use a 
 Gaussian window function,
\begin{equation}
W_{R_s}(k) = e^{- \frac{1}{2} (kR_s)^2} \;,
\end{equation}
in part to connect more closely with \citet{AloEarPea14}, as we use their
parameters as a starting point below.  One of the parameters of the
classification
is the smoothing scale $R_s$.  We
will show results from $R_s=4 h^{-1} Mpc$, studied in depth by \citet{AloEarPea14},
but note trends we saw for other
cases we explored, $R_s = (0.5, 1, 2, 8.9, 10, 19) \; h^{-1} Mpc$. 

Again, to make the notation clear,
because $D(a=1) = 1$, $\delta({\bf r})$ equals
the linearly evolved density perturbation today
and $\sigma$ corresponds to its commonly used definition as the
square root of the integral of the power spectrum linearly evolved to
the present day.  
Some other approaches include
  $D(a)$ in the overdensity, and thus in $\Phi$.  As the fluctuations
  $\sigma$ are
associated with the overdensity, in our definition they
 have a factor of $D(a)$ as
  well,
and so $\xi_{ij}$ remains
  time independent.  This is one reason for introducing $\sigma$ into
our definition of shear.  Other definitions of shear also sometimes take out the trace,
again not the case here.   When the shear in numerical simulations is considered, 
Eq.~\ref{eq:sheardef} is often used as well, but with the the fully evolved
nonlinear potential (determined by the evolved densities in the simulation),
rather than its linearly evolved counterpart used in the analytic work
in this note. 

The potential $\Phi$ can also be used to find the Zel'dovich approximation
to the displacement of initial positions at a later time.
In the Lagrangian approach to structure formation, 
the final position ${\bf x}$ of an initial mass
component with initial position ${\bf q}$ is related
via the displacement ${\bf \Psi}({\bf q},t)$:
\begin{equation}
{\bf x} = {\bf q} + {\bf \Psi}({\bf q},t).
\label{eq:zeldef}
\end{equation}
In the Zel'dovich approximation,
\begin{equation}
{\bf \Psi}^Z = D(a)\nabla_q \Phi \; .
\label{eq:zel}
\end{equation}
For the initial configuration, with coordinate ${\bf q}$,  multiplying the linear final time density
by $D(a)$ and taking $a$ small gives the initial density (which is linear because it is small,
$\delta_L({\bf r},a) = D(a) \delta({\bf r}) \sim 0$ as $a \to
0$ and ${\bf x} \to {\bf q}$).  

The shear can thus be thought of either as a property of the linearly
evolved final density perturbations, or as a property of positions at
early times (which then change, using Eq.~\ref{eq:zeldef}).  We will
discuss properties in both interpretations.

\subsection{Cosmic web classification}
The $3 \times 3$ shear matrix $\xi_{ij}$ has three eigenvalues,
traditionally ordered 
as $\lambda_1 \ge \lambda_2 \ge \lambda_3$.
Their sum, the trace of the shear, is proportional to the
density by Eq.~\ref{eq:delphi},
\begin{equation}  {\rm tr} \; \xi_{ij} =\sum_i \lambda_i (\bf r)
= \frac{1}{\sigma}\delta(\bf r).
\label{eq:deltalam}
\end{equation}

Halos, filaments, sheets and voids can then be defined as
regions where
(3,2,1,0) respectively of the $\lambda_i$ are above some threshold $\lambda_{\rm
  th}$.
In one way of estimating later structure formation,   
an initial region is expected to collapse at late times if it linearly
evolves to cross some overdensity threshold $\delta_c$.
As the linearly evolved overdensity $\delta ({\bf r}) = \sigma \sum_i \lambda_i({\bf r})$,
and regions classified here as halos might be expected to collapse at
late times (this is an assumption), this relation
suggests some minimum $\lambda_i$, i.e. nonzero 
$\lambda_{\rm th}$, for initial conditions.  Note that this
condition also means that the mean background, $\lambda_i = 0$, is also
classified as a void.  There are thus two scales in this
classification,
the smoothing $R_s$ and the threshold $\lambda_{\rm th}$.
Many different choices have been used in the literature.

The same eigenvalue classification is sometimes used for
dark matter simulations, but (as mentioned above) using 
the shear associated with
the fully evolved nonlinear potential, see,  e.g., \citet{Hah07,For09} and references therein.
For example, in dark matter nonlinear simulations evolved to redshift zero,
\citet{Hah07} found $R_s = 2 h^{-1}Mpc$ and $\lambda_{\rm th}=0$ 
separates structure into these different classes in a way that
matches visual estimates, while \citet{For09} considered $R_s = 0.88
h^{-1}Mpc$ and $\lambda_{\rm th} = 0, 0.2, 0.4., 1., 2.$ and found 
improvements for $\lambda_{\rm th}>0$ based
upon volume fractions and mass fractions.

One comparison between the above analytic and simulation approaches is
by \citet{AloEarPea14}.  In both approaches, they measured the mass
fraction and volume fraction in the four cosmic structures as a
function of $\lambda_{\rm th}$, and for two values of $R_s$.  They found
reasonable correspondence of volume and mass fractions for filaments,
sheets and voids between the analytically calculated linearly evolved
perturbations and the numerically evolved fully nonlinear
simulations.  Given this,
we looked for further characteristic properties of the cosmic web
in the analytic construction and report these properties here.
We focus for the most part on the parameters used in
\citet{AloEarPea14}.\footnote{There are other structure classifications using the
  eigenvalues, for instance based upon their sum, $\delta(\bf r)$ by
  Eq.~\ref{eq:delphi}, where each type of structure corresponds to $\delta({\bf r})$
 crossing some particular threshold
  $\delta_{\rm th}$ \citep{SheAbeMoShe06}.  The distributions of halos
  using $\lambda_{\rm th}$ vs $\delta_{\rm th}$ was done in
  \citet{LamSheDes09} where it was found that the $\lambda_{\rm th}$
  constraint worked well.
In analytic work inspiring the work here,
  \citet{PapShe13} considered a mixed constraint, using both
  $\lambda_{\rm th}$ and $\delta_{\rm th}$ to define halos and voids,
  and compared density anisotropies at large distances between the
  analytic (linearly evolved to final time) and simulated (nonlinearly
  evolved to final time) calculations. The mixed constraint makes the
  identification of filaments and sheets less direct.  We
  give properties using the \citet{PapShe13} constraint in Appendix \S\ref{sec:papsheex}.  Structures classified using eigenvalues of the Hessian of
  the density (e.g. \citet{Pog09}) and the Hessian of the velocities
  (e.g. \citet{Hof12}) have also been considered in great detail by
  many authors.}

We will use these definitions of halos, filaments, sheets and
voids to analytically calculate mean configurations of each,
for initial random Gaussian field
configurations with different choices of $R_s$ and $\lambda_{\rm th}$.
Unlike Press-Schechter theory where one changes the smoothing scale
to consider different mass halos, here the one smoothing scale produces a separation of
a full configuration into halos, filaments, sheets and voids, i.e. one
smoothing scale
completely classifies an entire map.  Our
mean objects correspond to one such specific smoothing scale and threshold.

\subsection{Average shear eigenvalues}
The analytic definition of halos, filaments, sheets and voids in
terms of $\lambda_{\rm th}$ allows the calculation of
 volume and mass fractions in 
each type of object and 
the average value of each of the eigenvalues
$\lambda_1,\lambda_2,\lambda_3$
in the Gaussian random field initial conditions.

The probability distribution for the ordered eigenvalues $\lambda_1\ge\lambda_2\ge\lambda_3$ of a
Gaussian random field is \citep{Dor70}
\begin{equation}
{p}({\lambda}_1,{\lambda}_2,{\lambda}_3) = \frac{3375}{8 \sqrt{5} \pi}
e^{-3 I_1^2- \frac{15}{2} I_2} ({\lambda}_1-{\lambda}_2)({\lambda}_1-{\lambda}_3)({\lambda}_2-{\lambda}_3)
\end{equation} 
where 
$I_1 = {\lambda}_1 +{\lambda}_2 +{\lambda}_3$ and
$I_2 = {\lambda}_1 {\lambda}_2 + {\lambda}_2 {\lambda}_3 + {\lambda}_1
{\lambda}_3$.  Note that our $\lambda_i$ differ by a factor of
$\sigma$ from that in \citet{Dor70} because we include $\sigma$ in our
definition of shear, Eq.~\ref{eq:sheardef}.
Expressions for pairs of eigenvalues or individual eigenvalues are
found in \citet{LeeSha98} and another interesting rewriting is in
\citet{AloEarPea14}. 

The probability of the eigenvalues restricted by some condition $C$,
given for example
in terms of
$\lambda_{\rm th}$, is given by
\begin{equation}
P(C) = \int_C d^3 \lambda p(\lambda_1,\lambda_2,\lambda_3) \; .
\label{eq:pccalc}
\end{equation}
One case of interest for us is
where the shear in the region $C$ obeys the constraints for a halo, filament, sheet
or void in terms of $\lambda_{\rm th}$. 
The mean eigenvalues for this particular constraint $C$, which we will denote by
$\bar{\lambda}_{i,CA}$, are
\begin{equation}
\begin{array}{l} {\rm halo} \\ {\rm filament} \\ {\rm sheet} \\ {\rm void}
\end{array}
\left.
\begin{array}{l}
\int_{\lambda_{\rm th}}^\infty d \lambda_1 \int_{\lambda_{\rm th}}^{\lambda_1} d
\lambda_2 \int_{\lambda_{\rm th}}^{\lambda_2} d \lambda_3  \\
\int_{\lambda_{\rm th}}^\infty d \lambda_1 \int_{\lambda_{\rm th}}^{\lambda_1} d
\lambda_2 \int_{-\infty}^{\lambda_{\rm th}} d \lambda_3  \\
\int_{\lambda_{\rm th}}^\infty d \lambda_1 \int_{-\infty}^{\lambda_{\rm th}} d
\lambda_2 \int_{-\infty}^{\lambda_2} d \lambda_3  \\
\int_{-\infty}^{\lambda_{\rm th}} d \lambda_1 \int_{-\infty}^{\lambda_1} d
\lambda_2 \int_{-\infty}^{\lambda_2} d \lambda_3  
\end{array} \right\}
\times p(\lambda_1,\lambda_2, \lambda_3) \frac{ \lambda_i}{P(C)} \; .
\label{eq:dorav}
\end{equation}
In this example, for each structure the corresponding normalization
$P(C)$, Eq.~\ref{eq:pccalc}, has the same integration limits as the
the corresponding integral over
$p(\lambda_1,\lambda_2, \lambda_3) { \lambda_i}$ (it changes for
different structures).  For example, for a halo,
\begin{equation}
\bar{\lambda}^{\rm halo}_{i,CA} \equiv
\frac{\int_{\lambda_{\rm th}}^\infty d \lambda_1 \int_{\lambda_{\rm th}}^{\lambda_1} d
\lambda_2 \int_{\lambda_{\rm th}}^{\lambda_2} d \lambda_3
p(\lambda_1,\lambda_2 \lambda_3) \lambda_i }
{\int_{\lambda_{\rm th}}^\infty d \lambda_1 \int_{\lambda_{\rm th}}^{\lambda_1} d
\lambda_2 \int_{\lambda_{\rm th}}^{\lambda_2} d \lambda_3
p(\lambda_1,\lambda_2 \lambda_3) }
\label{eq:haloCA}
\end{equation}
We use the notation
$\bar{\lambda}_{i,CA}$ to indicate these constrained averages.  It
depends both upon $\lambda_{\rm th}$ and the type of structure (halo,
filament, sheet or void).
In practice, for any given $\lambda_{\rm th}$ we integrated these
expressions numerically using mathematica.

Another constraint goes beyond restricting eigenvalues in terms
$\lambda_{\rm th}$  for each structure, requiring the $\Phi$ from
which they come to also correspond to
extremal values of the potential ($\partial_i \Phi = 0$).  
In this case the shear 
obeys conditions to be a halo, filament, sheet or void in terms of
$\lambda_{\rm th}$, and its
potential also obeys
$\partial_i \Phi=0$.  The measure for the volume average of points
obeying this constraint is found in
\citet{BBKS}, starting with the probability
distribution for the random Gaussian fields
$\Phi,\partial_i\Phi,\xi_{ij}$ times the constraint on $\partial_i \Phi$.  
They were interested in peaks and minima in particular, as their main interest
was for halos and voids (they also had a restriction on $\Phi$,
which we do not impose here, and thus it drops out of the expressions).
Their initial measure for integrating the probability density over
volume, shear and $\partial_i \Phi$ becomes an
 integral over the eigenvalues similar to that in
Eq.~\ref{eq:dorav}, except with an additional measure factor 
$|\lambda_1 \lambda_2 \lambda_3|$ (from changing
variables in the constraint $\delta_D(\partial_i\Phi({\bf r}))$ from
$\partial_i \Phi$ to position
and integrating over position). 
\begin{equation}
\begin{array}{l}
\int d^3 r \int d^6 \xi \int d^3 \partial \Phi \delta({\bf r-r_p})
p(\xi_{ij}, \partial_i \Phi)\\ =
\int d^3 \lambda \int d^3 \partial \Phi \delta({\partial \Phi})
p({\lambda}_1,{\lambda}_2,{\lambda}_3) |\lambda_1 \lambda_2 \lambda_3|
\; .
\end{array}
\label{eq:extrfactor}
\end{equation}
Here the integral again is over the values of $\lambda_i$ obeying the
constraint to be a halo, filament, sheet or void.
We will call the corresponding average shear values $\bar{\lambda}_{i,{\rm
    extr}}$.
So, for example, for a halo, one has:
\begin{equation}
\bar{\lambda}^{\rm halo}_{i,{\rm extr}} \equiv
\frac{\int_{\lambda_{\rm th}}^\infty d \lambda_1 \int_{\lambda_{\rm th}}^{\lambda_1} d
\lambda_2 \int_{\lambda_{\rm th}}^{\lambda_2} d \lambda_3
p(\lambda_1,\lambda_2 \lambda_3) |\lambda_1 \lambda_2 \lambda_3|\lambda_i }
{\int_{\lambda_{\rm th}}^\infty d \lambda_1 \int_{\lambda_{\rm th}}^{\lambda_1} d
\lambda_2 \int_{\lambda_{\rm th}}^{\lambda_2} d \lambda_3
p(\lambda_1,\lambda_2 \lambda_3) |\lambda_1 \lambda_2 \lambda_3| }
\label{eq:haloextr}
\end{equation}

\begin{figure}
\begin{center}
\resizebox{3.in}{!}{\includegraphics{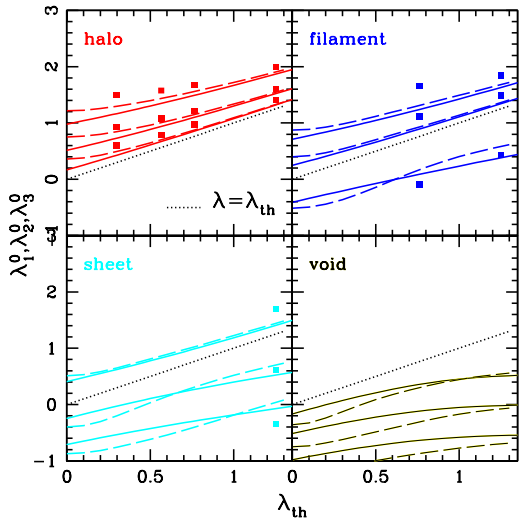}}
\end{center}
\caption{
Average values of 3 constrained eigenvalues as a function of
threshold $\lambda_{\rm th}$ for each kind of structure.  The solid
line is $\bar{\lambda}_{i,CA}$ and the dashed line is
$\bar{\lambda}_{i,{\rm extr}}$, the
dotted line is the threshold $\lambda_{\rm th}$.  Eigenvalues for
halos (red), filaments (blue), sheets (cyan), and voids (yellow, with
black central lines for clarity) 
are shown at upper left, upper right, lower left, and lower
right.
These curves are also the central values $\lambda^0_i$ for two of the 
three mean map constructions in \S\ref{sec:choices},
$\lambda^0_i= \bar{\lambda}_{i,CA}$ for choice 2 and $\lambda^0_i =
\bar{\lambda}_{i,{\rm extr}}$ for choice 3.
The filled squares are for choice
1,
where $\lambda^0_i$ is fixed to make the average shear
$\bar{\lambda}_i$ within the central
object satisfies $\bar{\lambda}_i = \bar{\lambda}_{i,CA}$.
The mean maps hereon use the same colors (for voids, yellow alone) for each kind of structure.
}
\label{fig:lamcens}  
\end{figure}

These two kinds of average shear, $\bar{\lambda}_{i,CA}$ and
$\bar{\lambda}_{i,{\rm extr}}$, are shown in
Fig.~\ref{fig:lamcens} below for the 4 kinds of objects, as the dashed
and solid lines, and are quite similar.

We will consider one other shear average below, that of the mean objects
we construct.  The eigenvalues of the average shear within these mean
objects will be called $\bar{\lambda}_i$.

\subsection{Mean shear maps: defining objects and characteristic properties}
One can go beyond 
the analytic volume fractions, mass fractions and the constrained shear
eigenvalue averages for each structure type by using the power spectrum.
The power spectrum can be used to calculate mean spatial properties of
the shear around any given central shear configuration.  We make maps of
 these mean shear values around a given central shear value, and study properties of
regions within them, in particular those regions sharing the same cosmic web
classification as the central point.

We use the method of \citet{PapShe13}, following \citet{BBKS}, 
to find the mean values of
 $\xi_{ij}({\bf r})$ (and thus $\lambda_i({\bf r})$) everywhere around 
a given configuration ${\xi}_{ij}(0) \equiv {\xi}_{ij}$ at the origin.  Their 
calculation, for our case of interest, is summarized in Appendix \S\ref{sec:shearcalc}.
Denoting the constraint at the origin as $C_0$,
we thus get a map $\langle \xi_{ij}({\bf r})|C_0 \rangle$, a mean shear
at every point.  This shear map depends not only on $C_0$ as 
explicitly indicated, but also
upon the linear power spectrum and smoothing.  The shear eigenvalues
at the origin corresponding to the origin constraint $C_0$ 
are denoted as $\lambda_i^0$.\footnote{For 
simplicity we will always take a
coordinate system such that the shear tensor is diagonalized at the
origin,
with the largest value of shear along $x$, second along
$y$ and smallest along $z$.}

Once the mean shear $\langle \xi_{ij}({\bf r})|C_0
\rangle$ is known everywhere around a given central
configuration, a choice of $\lambda_{\rm
  th}$ separates the mean map
into halos, filaments, sheets or voids.  If the origin
constraint $C_0$ corresponds to a halo, the nearest points are usually
also halo points, similarly for filaments, sheets and voids. 
We consider the central objects in the mean maps in the most detail.
The central object in a mean map is the region sharing
the same cosmic web classification (e.g. filament) as the origin and which are
continuously connected to the origin via regions with this same classification.
Far away from the origin, 
the correlations and thus the mean shear go to zero
and the mean configuration becomes the mean background (again,
this is a void region for
$\lambda_{\rm th}>0$).  In between the origin and mean background,
depending upon parameters, several different types of structures may
occur.  For instance
mean halos tend to have filaments and then sheets around them
before reaching the mean 
background.\footnote{See the nice analytic toy model in
  \citet{AloEarPea14} for an intuitive example of structure around a
  halo.  In terms of the shear eigenvalues, configurations
going from all three eigenvalues above threshold to 
no eigenvalues above threshold would
generally be expected to pass through configurations 
with two or one of the eigenvalues 
above threshold.} 
Similarly, filaments
tend to have sheets around them (and sometimes halos within, towards their
ends).  Voids do not always have a clear boundary because for $\lambda_{\rm
  th}>0$ they also can include the mean background.

Several quantities can be associated with these resulting mean map objects,
such as 
volume, shape, and average values of shear 
$\bar{\lambda}_i$\footnote{The average eigenvalues are calculated by
taking all points classified as belonging to the structure by their
eigenvalues,
averaging their shear over the volume of the object, 
and then finding the eigenvalues of this average shear.}.  All of these depend upon 
$C_0$, $\lambda_{\rm th}$ and $R_s$. 

We report the volume of these mean objects in terms of the  Lagrangian volume for
a halo associated with the same window function and smoothing scale $R_s$,
$V_{\rm halo \; Lag}$.  This Lagrangian volume is found by
by integrating the background average density at early times (as $\delta_L({\bf r},a)
\to 0$) against
the window function to give $M(R_s)$ and then converting to $V_{\rm
  halo \; Lag}=M(R_s)/\rho_b$.  For our Gaussian window function,
$V_{\rm halo \; Lag} = (2 \pi)^{3/2} R_s^3$.
The correspondence between shear and initial Gaussian perturbations
means that one can view
the volumes of objects in our maps as Lagrangian volumes as well,
as their volume also directly gives their mass at early times.

For one choice of shear at the origin,
$\lambda_i^0=\bar{\lambda}_{i,{\rm extr}}$,
boundary conditions for the Zel'dovich displacement at the origin are
also specified, as $\partial_i \Phi = 0$.  In this case we also 
calculate the mean linearly extrapolated Lagrangian displacement at late times given by
Eq.~\ref{eq:zel}.  The mean displacement of the edges of the mean objects gives
one way of estimating (using linear evolution) their final size.

An additional measurement we make is of the average of the
linearly extrapolated Eulerian overdensity at final
time in these mean objects, given by $\langle
\delta \rangle = \sigma \sum_i \bar{\lambda}_i$.  In simulations, this
has been measured by running the clock backwards, starting
with
collapsed halo regions in nonlinear simulations.  Particles in the
final halo are tagged and then traced back in time, and
the average of the shears of these particles in the initial conditions is used to
calculate the linearly extrapolated Eulerian overdensity at final times.
\citet{DesTorShe13} found for lower final
halo masses that on average $\delta_{\rm extrap}  \sim \delta_c (1+0.2
\sigma(R_s))$, and \citet{DalWhiBonShi08} found similar trends for higher
mass halos.\footnote{The definition of halo at final times and the
  cosmological parameters differed in these studies, and the measured
  relations had a lot of scatter, so these numbers are only
  approximate, another similar estimate was found in \citet{EliLudPor12}.  See \S \ref{sec:collapse} for more discussion of
relations between collapsed halos and halos in this shear based classification.}   We quote overdensities in terms of this reference $\delta_{\rm
  extrap}$ below.  

To summarize, for central halo, filament and
sheet regions in mean maps we will measure  
volume,
shape (axes lengths), average
shear eigenvalues within,
 $\bar{\lambda}_i$, linearly extrapolated average overdensity$\langle
\delta \rangle$, and in some cases mean Zel'dovich linearly
extrapolated displacement ${\bf \Psi}^Z$.  
Most of our focus is on the objects which include the origin (objects
not connected to the origin will be discussed briefly in 
Appendix \S\ref{sec:subtle}).

We show examples below of cross sections of these regions along the 3
pairs of spatial axes and
of the three shear eigenvalues and
linearly extrapolated overdensity along the 3 spatial axes.

\section{Shear Map Examples}
\label{sec:examp}

The mean maps for a given cosmology are determined by
the choice of central shear $\langle \xi_{ij} |C_0\rangle=
(\lambda^0_1,\lambda^0_2,\lambda^0_3)$, the threshold defining the
separation into different structures,
$\lambda_{\rm th}$, the smoothing $R_s$, the window function and the
power spectrum.  We hold the window function (Gaussian) and power
spectrum fixed and show results for $R_s = 4 h^{-1} Mpc$
(commenting on
changes for other values).
The threshold $\lambda_{\rm th}$
and central shears $C_0$ vary as indicated.

\subsection{Choices for shear at origin, $\lambda^0_i$, of mean maps}
\label{sec:choices}
To construct a mean map, a choice of shear at the origin, $C_0$, is required.
Mean maps can be constructed for any choice of
shear at the origin $C_0$, but for a given type of structure and 
$\lambda_{\rm th}$, we found three choices of particular 
interest, all motivated by the analytic
calculations of average shear given some constraint.  Two are
associated with $\bar{\lambda}_{i,CA}$, found via Eq.~\ref{eq:dorav},
and
one with $\bar{\lambda}_{i,{\rm
    extr}}$, which includes
Eq.~\ref{eq:extrfactor} in the calculation of the average shear
eigenvalues.  
They all yield a value for $\lambda^0_i$, the shear at the origin as follows
(implicitly for choice 1, explicitly for choices 2,3).

\begin{figure*}
\begin{center}
\resizebox{3.in}{!}{\includegraphics{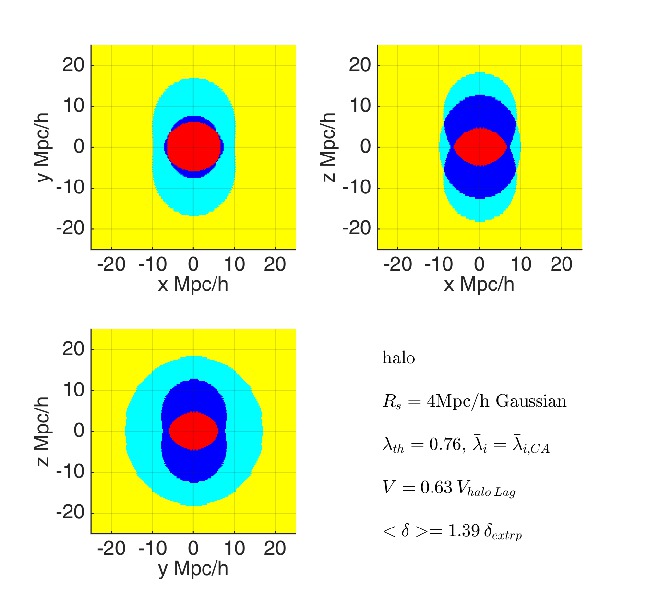}}
\resizebox{3.in}{!}{\includegraphics{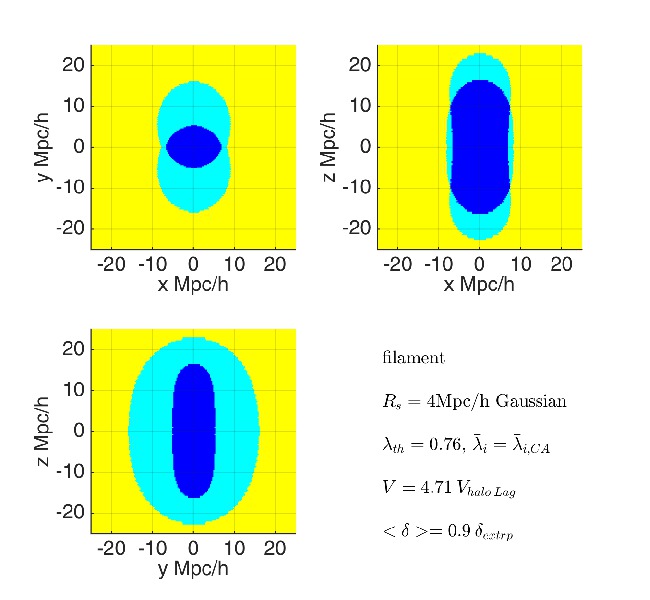}}
\resizebox{3.in}{!}{\includegraphics{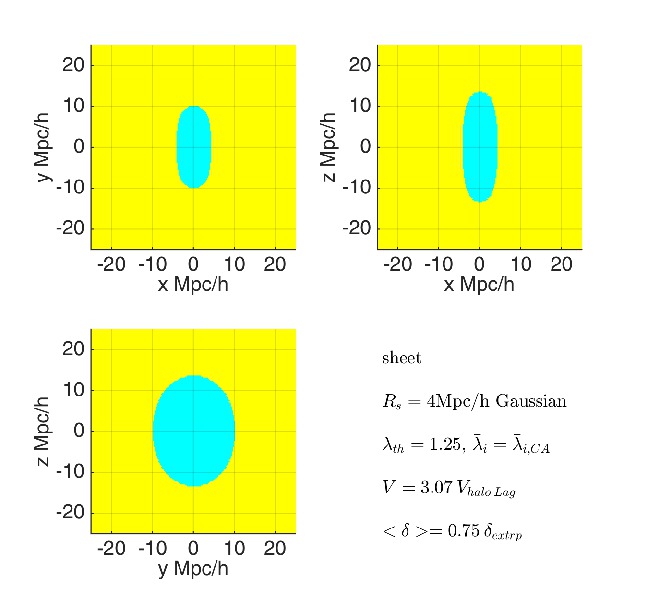}}
\end{center}
\caption{Cross sections of a central mean halo (top left), filament
  (top right) and sheet (bottom) along three pairs of axes.  For
each central object, $\lambda^0_i$ is calculated for the chosen
$\lambda_{\rm th}$ (listed in each plot) using choice 1: 
the average eigenvalues $\bar{\lambda}_i =
\bar{\lambda}_{i,CA}$ for the corresponding type of structure (halo,
filament or sheet).  At top left, 
the mean halo (red)
is surrounded by a filament region (dark blue), then a sheet region (cyan)
and finally a void (yellow, note mean background is also a ``void'').  
The smoothing $R_s = 4 h^{-1} Mpc$ for all three,  the threshold
$\lambda_{\rm th}$ is 
as indicated.  
The largest eigenvalue is along $x$.  For the central object in each,
volume and linearly extrapolated overdensity $\langle \delta \rangle$
are also listed.  )No mean void is illustrated, as shear corresponding to a void
at the origin can smoothly transform to the mean background, which is
also a void.  It is thus difficult to define the
 volume of the void and find a central value using this choice, choice 1, for $\lambda^0_i$.)
 }
\label{fig:aep}  
\end{figure*}

\begin{figure*}
\begin{center}
\resizebox{3.in}{!}{\includegraphics{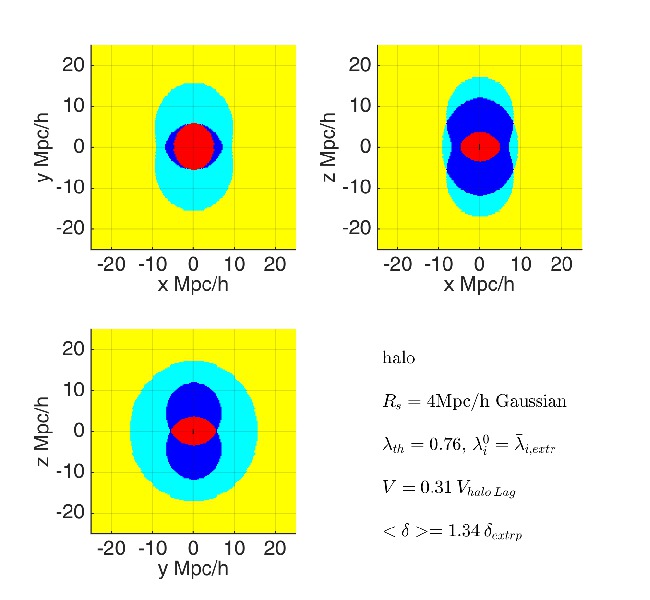}}
\resizebox{3.in}{!}{\includegraphics{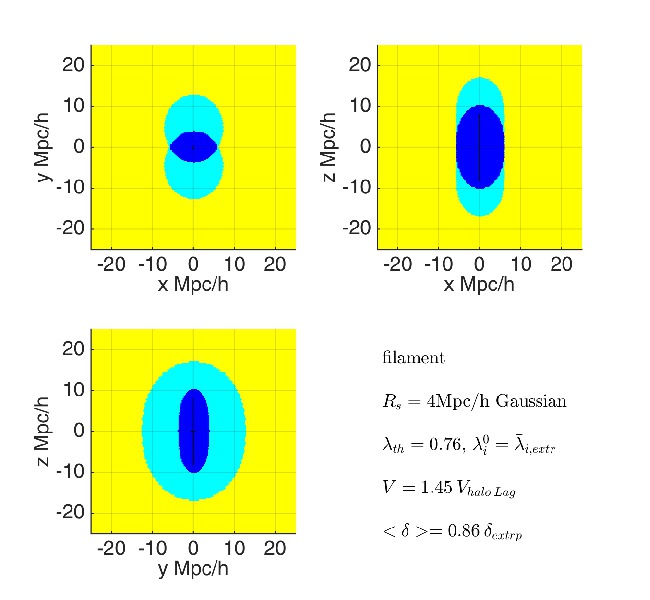}}
\resizebox{3.in}{!}{\includegraphics{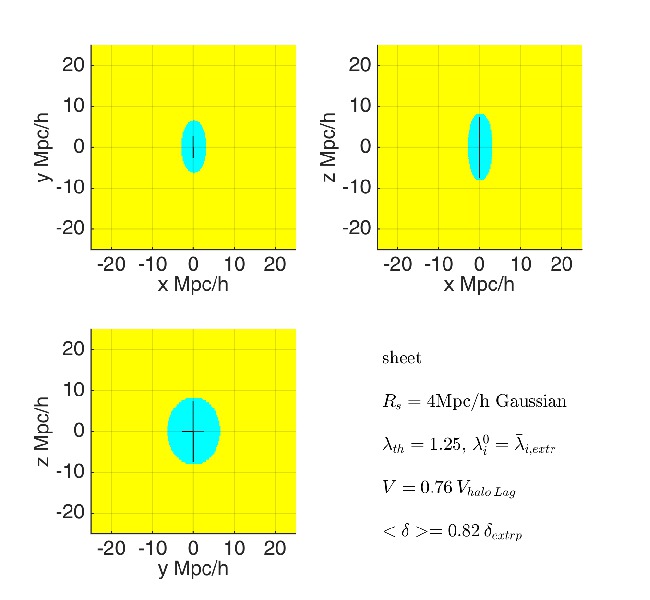}}
\end{center}
\caption{Cross sections analogous  
Fig.~\ref{fig:aep}, but for central shear $\lambda^0_i =
\bar{\lambda}_{i,{\rm extr}}$ (case 3 in \S\ref{sec:choices}).  Colors,
$R_s$, and $\lambda_{\rm th}$ as 
in Fig.~\ref{fig:aep}. 
In addition, initial edge positions displaced using the mean Zel'dovich
approximation,
Eq.~\ref{eq:zel},
are connected across the origin via black lines (if they do not go beyond the
origin). No mean void is shown as the central
configuration smoothly transitions to the mean background, which is also a void.
 }
\label{fig:bbks}  
\end{figure*}

\begin{figure*}
\begin{center}
\resizebox{3.in}{!}{\includegraphics{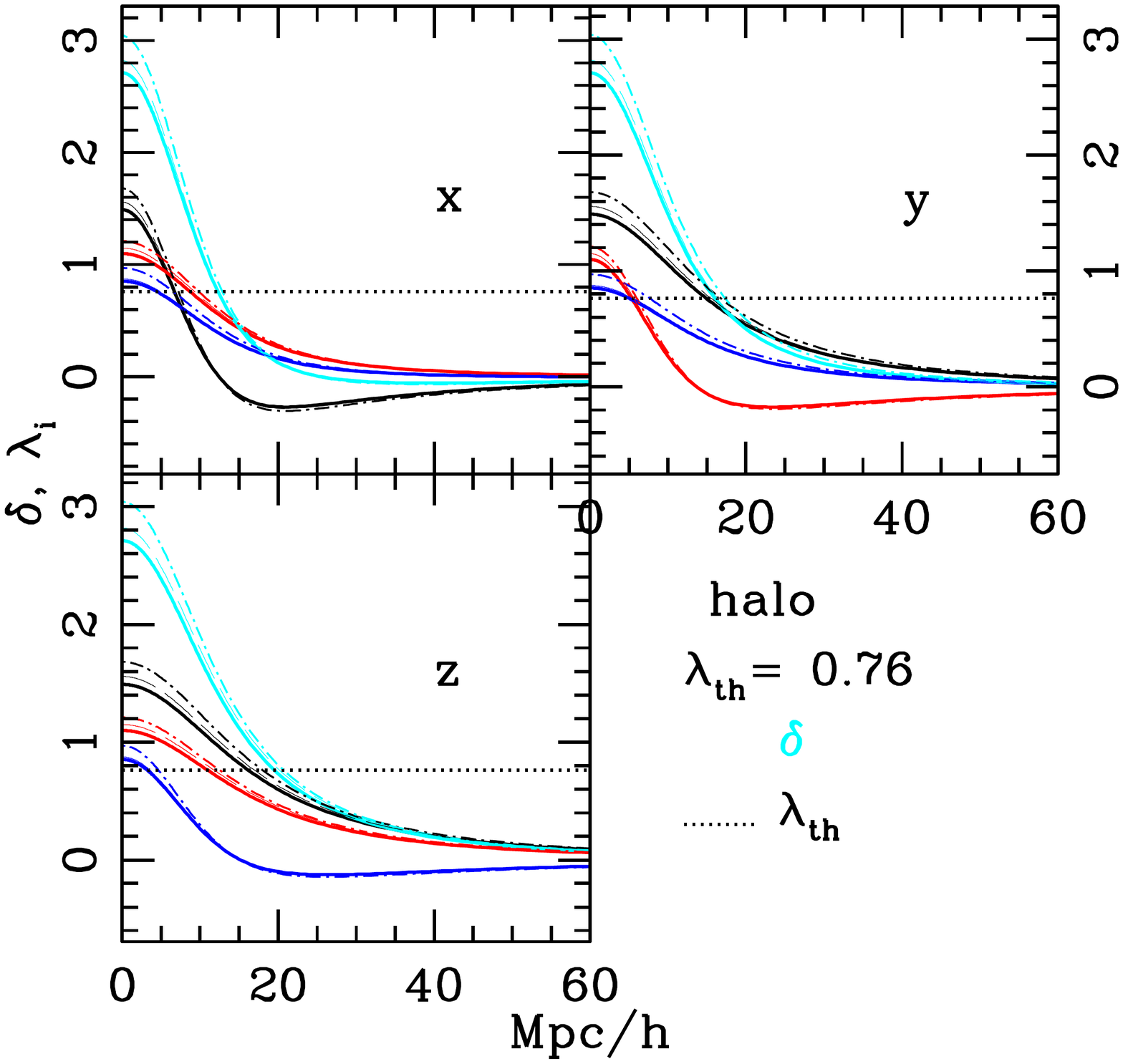}}
\resizebox{3.in}{!}{\includegraphics{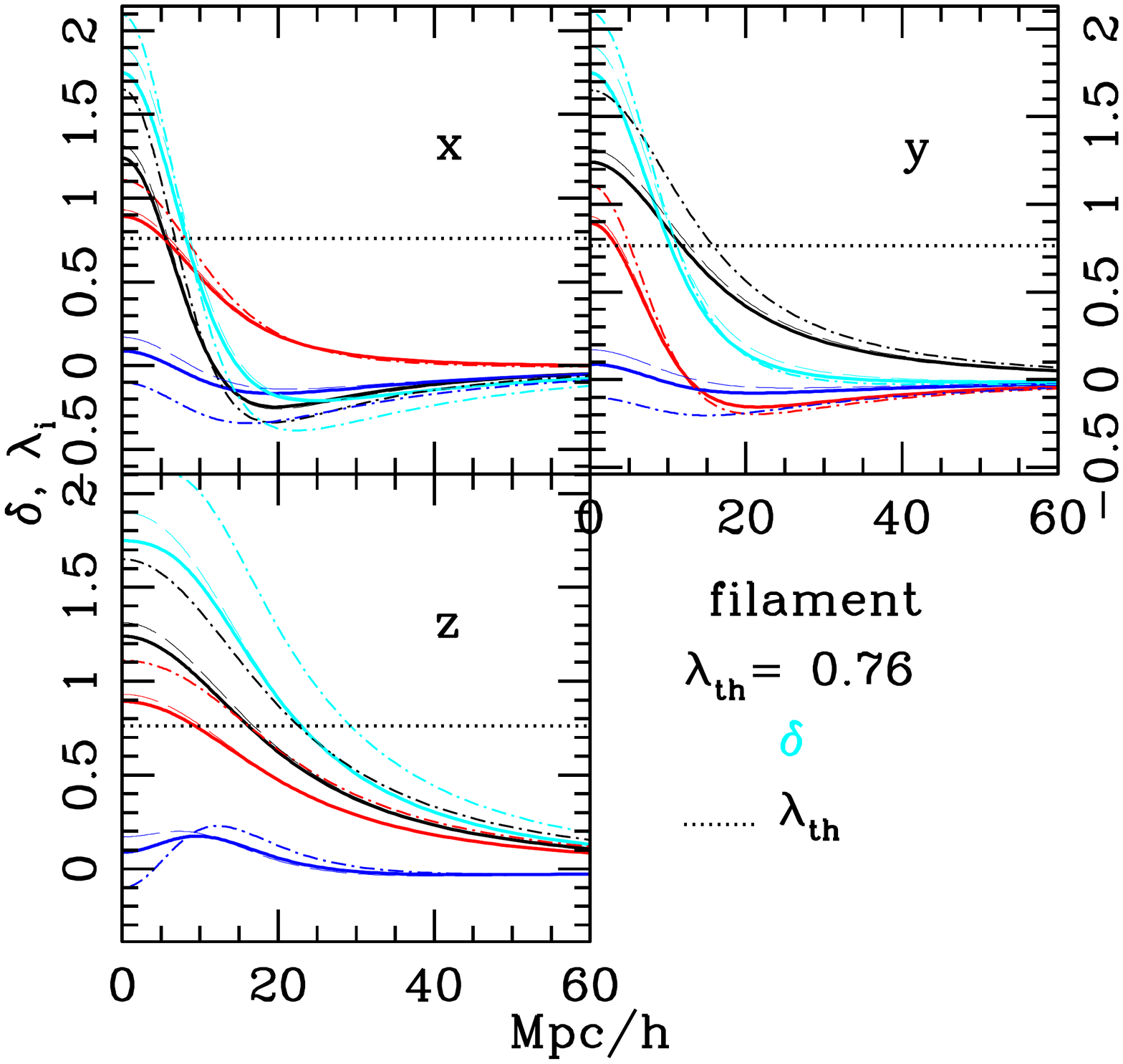}}
\resizebox{3.in}{!}{\includegraphics{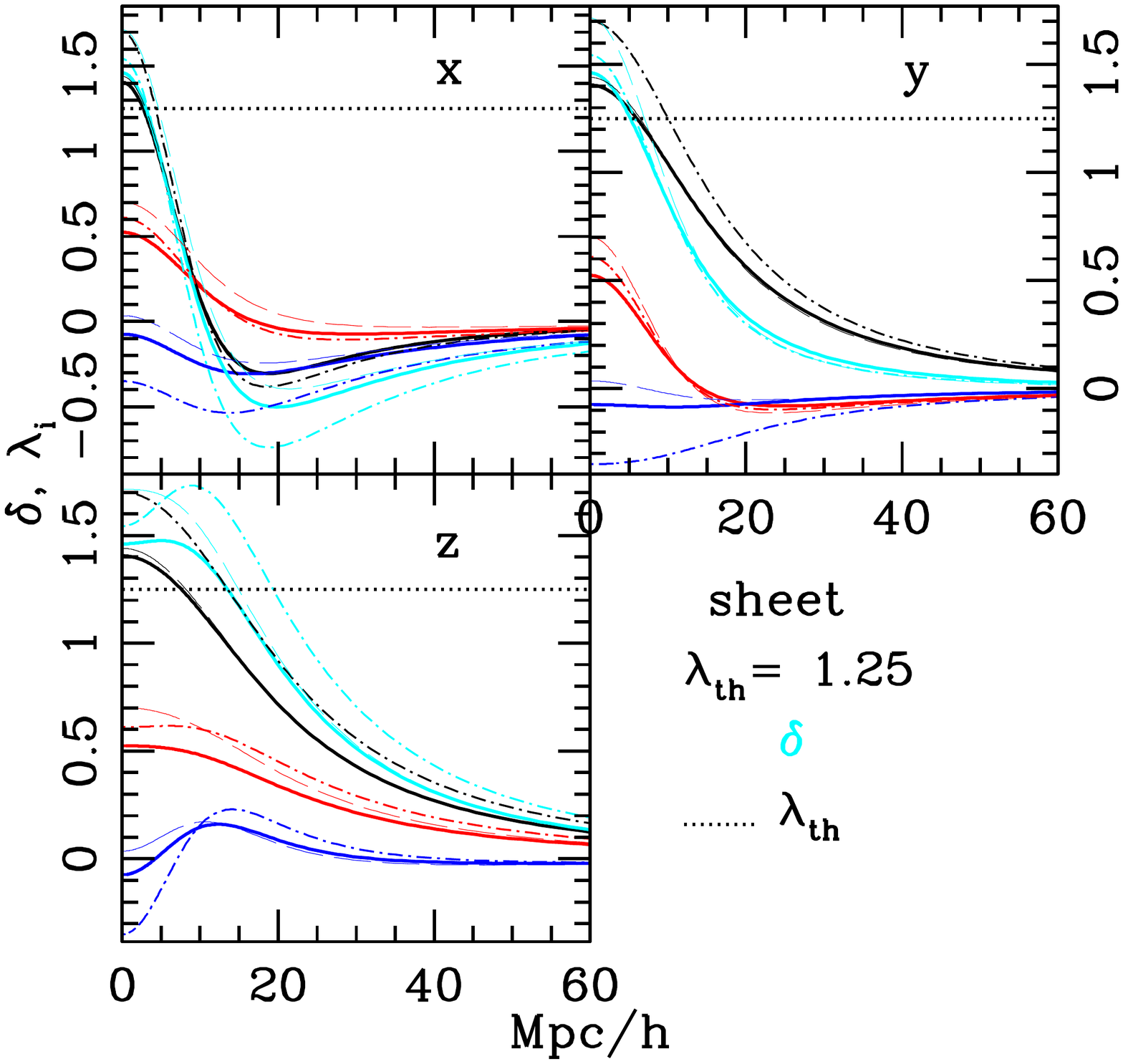}}
\resizebox{3.in}{!}{\includegraphics{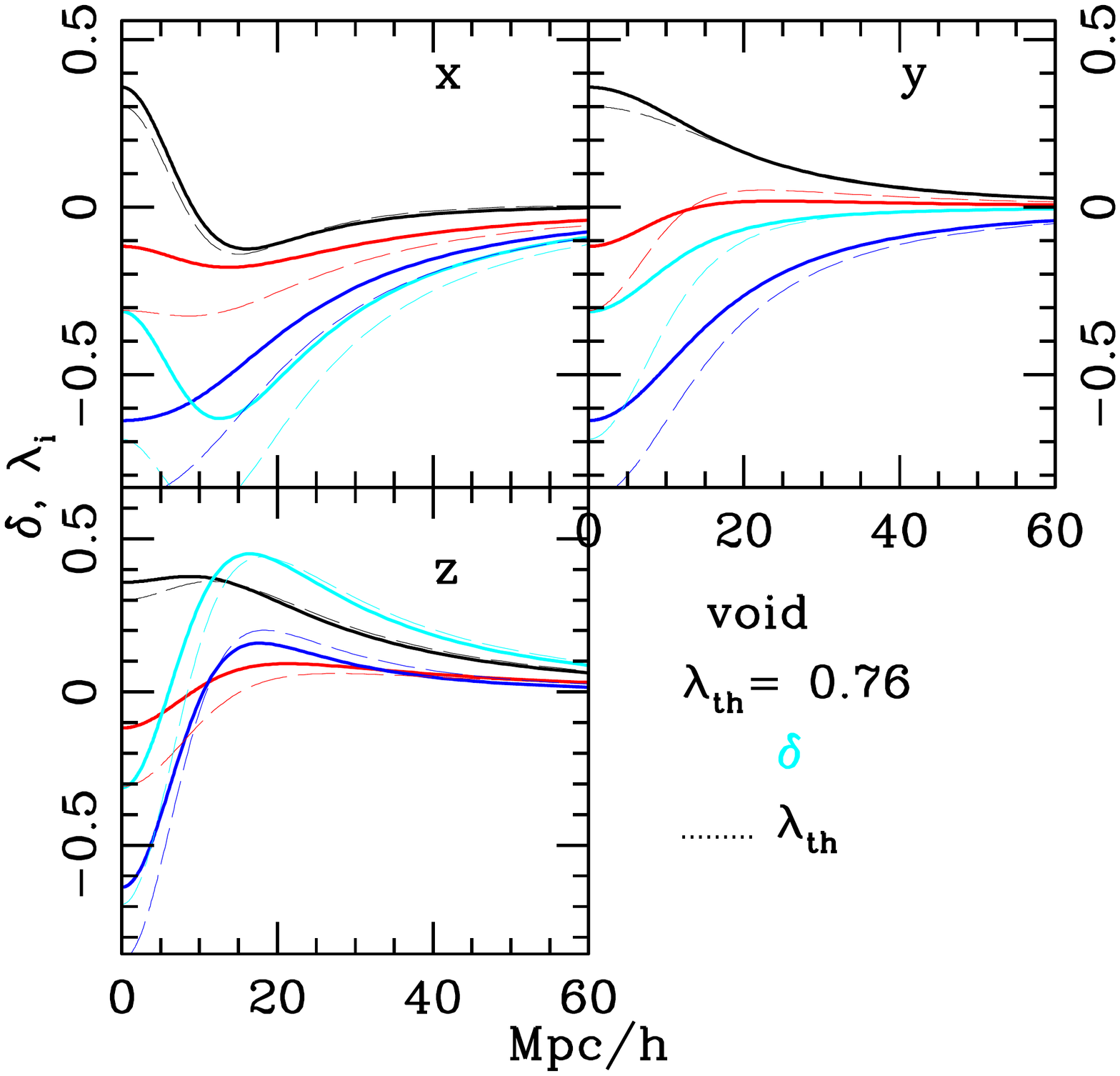}}
\end{center}
\caption{
Shear eigenvalue profiles $ \lambda_i (r)$
  (black, blue, red)
 and linearly  extrapolated overdensity $\delta$ (cyan) along
 $x$,$y$ and $z$ axes, for 
a halo (upper left panels), filament (upper right panels), sheet
(lower left panels) and void (lower right panels).  For the
$\lambda_{\rm th}$ as in Fig.~\ref{fig:aep},\ref{fig:bbks} (and
$\lambda_{\rm th} = 0.762$ for the void), profiles for
our three cases of central shear are shown: 
(1) $\bar{\lambda}_i
=\bar{\lambda}_{i,CA}$ (medium weight,
dot-dashed lines),
(2) $\lambda^0_i =
\bar{\lambda}_{i,CA}$ (heavier
solid lines) and (3) $\lambda^0_i = \bar{\lambda}_{i,{\rm
    extr}}$ (lighter dashed
lines). 
The straight black dotted line is the threshold $\lambda_{\rm th}$
separating out the different structures (not visible for 
the void example at lower right).
}
\label{fig:profiles}  
\end{figure*}

\begin{description}
\item[1.] $\bar{\lambda}_i=\bar{\lambda}_{i,CA}$ :\\
Origin shear choice 1 fixes $\lambda^0_i$ implicitly, by requiring that
the average $\lambda_i$ within a particular mean object in
the map, $\bar{\lambda}_i$, corresponds to
$\bar{\lambda}_{i,CA}$ for the corresponding type of structure.  That is, the
mean object with shear $\lambda^0_i$ at the origin has average shear corresponding to the
average analytic shear, Eq.~\ref{eq:dorav} for that kind of structure.
For example, for a halo in the mean map, 
$\bar{\lambda}^{halo}_i=\bar{\lambda}^{\rm halo}_{i,CA}$, where the
left hand side is the average over the values in the mean map, and
similarly for the corresponding filament (using $\bar{\lambda}^{\rm
  fil}_{i,CA}$, etc.).
In practice, we change the central value of shear until the mean
shears within the halo region centered at the origin (which changes in
size and shape as the central shear changes), obeys
Eq.~\ref{eq:haloCA}.
One possible 
interpretation is that this gives
an ``average'' such object (e.g. halo), in a smoothed
simulation.
Practically, we could not always find solutions $\lambda^0_i$
satisfying 
the constraint $\bar{\lambda}_i = \bar{\lambda}_{i,CA}$ for all
$\lambda_{\rm th}$ and all
structures, in addition, other
complications in this implicit definition can arise
(Appendix \S\ref{sec:subtle}).  Our examples below are in the regime
where
the associated corrections from these possible complications seem small enough to be ignored and where
we could find solutions.
\\

\item[2.]
  ${\lambda}^0_i=\bar{\lambda}_{i,CA}$
  :\\
Origin shear choice 2 sets the central shear $\lambda^0_i$ equal to the average
shear $\bar{\lambda}_{i,CA}$ in Eq.~\ref{eq:dorav} (similar to
\citet{PapShe13}, who used it for long distance properties at final
times and whose work inspired this study).
This means putting the average shear for a certain structure (for
instance a halo)
at the origin and then measuring properties of the object 
made up of the enclosing region of points sharing the same cosmic web 
classification.  In more detail, for a halo,
the origin would have shear eigenvalues
$\lambda^0_i=\bar{\lambda}^{\rm halo}_{i,CA}$.
Although slightly more complicated to interpret, these maps 
are much easier to construct, as the central shear is explicitly and
immediately calculable.
Mean properties around such a point might
be interpreted as mean properties around an average point in 
the corresponding structures in a large map, rather than properties
around
the central point of a structure (i.e. the peak of a halo does not
have
average halo properties).\footnote{
The mean objects constructed this way might have sizes more closely
related to, e.g.,
how far (on average) one might have to go, from an
average point for a given structure, before the object ends.
  The approximate radial position of such an average point in the mean
  maps above for $\bar{\lambda}_i = \bar{\lambda}_{i,CA}$ can be read
  off from Fig.~\ref{fig:profiles}.}
\\

\item[3.] ${\lambda}^0_i=\bar{\lambda}_{i,extr}$ :\\
Origin shear choice 3 is the volume average of 
shears obeying some particular cosmic web constraint where in addition $\Phi$ is at an extrema, 
i.e. where $\partial_i \Phi=0$ for all $i$.  This is reasonable
to expect for a center of a halo or void (minimum or peak) and we can
also consider this condition for the center of a filament or sheet.
These $\lambda_i^0 = \bar{\lambda}_{i,{\rm extr}}$ are calculated with
the extra measure
factor derived from Eq.~\ref{eq:extrfactor}.  Intuitively, the mean
objects corresponding to these central shear values, at least for
halos and voids, might correspond to stacking objects in a map,
centered on their extremal points (which is often the point used to
define the center of such objects).

\end{description}

In Fig.~\ref{fig:lamcens} we show the central values for halos,
filaments, sheets and voids for these three constructions, as a
function of $\lambda_{\rm th}$.  As central value 1 required
implicit solving of an equation, only a few values are shown, as points (and
again, we could not find solutions for all $\lambda_{\rm th}$ in this
case, and the construction is ill-defined for voids).  
The solid line is for case 2, $\lambda^0_i = \bar{\lambda}_{i,CA}$,
the integral in Eq.~\ref{eq:dorav}, the dashed line is for case 3,
the volume average over extremal points, 
$\lambda^0_i = \bar{\lambda}_{i,extr}$.  The central values are very
similar for the three constructions.

\subsection{Mean object cross sections}
With these central values we can calculate the mean shear around the origin,
and classify each region as halo, filament, sheet, or void.
In Fig.~\ref{fig:aep}, we show example cross sections of regions obeying 
constraint 1 for the central object, i.e. where $\bar{\lambda}_i$ within the central object is
approximately equal to the corresponding $\bar{\lambda}_{i,CA}$, along
the three pairs of axes.  We use colors to distinguish between structures:
red for halos, dark blue for filaments, cyan for
sheets and yellow for voids.\footnote{For 
the central halo and filament, we use parameters 
studied in detail in
\citet{AloEarPea14},
smoothing $R_s=4 h^{-1} Mpc$, and threshold $\lambda_{\rm th}=0.762$.  
They chose $\lambda_{\rm th}$
by trying to get
similar analytic volumes in each of the different types of
structures.
For the sheet shown we changed $\lambda_{\rm th}$ to 1.25 to avoid
subtleties discussed in Appendix \S\ref{sec:subtle}.} 
The mean background ($\lambda_i = 0$) is
also a void (yellow) for any $\lambda_{\rm th}>0$. 
 (As voids comprise the eventual mean background into
which these structures are found, and the majority of the volume,
as noted earlier it was not clear how to impose $\bar{\lambda}_i =
\bar{\lambda}_{i,CA}$ for their case.
Thus no void example is shown.)
As anticipated, the mean central halo (at top left, Fig.~\ref{fig:aep}) is
surrounded by a filament, then a sheet, and then eventually
reaches mean density (which is a void), similarly the mean filament
(at top right, Fig.~\ref{fig:aep}) is
surrounded by a sheet and then void and mean density, and so on.
The mean filament does not extend between halos for this value of
$\lambda_{\rm th}$ although as $\lambda_{\rm th}$ is lowered halos do
eventually appear in the mean filament endpoints.
The mean halo has comparable size in all three directions, the filament is extended in
one direction compared to the other two, and the sheet has one
direction much smaller than the other two, as expected.  The larger
directions are a few times larger than the smoothing scale $R_s$.

Fig.~\ref{fig:aep} also lists the mean extrapolated
overdensity and volume for each
central object.  (By construction, for choice of origin 1,
$\langle \delta \rangle= \sigma \sum_i \bar{\lambda}_{i,CA}$.)
For the $\lambda_{\rm th}$ and $R_s$ here (0.762 and $4 h^{-1} Mpc$
respectively), the halo volume is slightly smaller than expected for
the Lagrangian region associated with a halo of mass $M(R_s)$, and
the extrapolated density is higher,
$V\sim 0.6 V_{\rm halo \; Lag}, \delta \sim 1.4 \delta_{\rm extrap}$.\footnote{
This doesn't always happen for
other parameters.  For instance, for the other set of $R_s, \lambda_{\rm th}$ explored in
detail by \citet{AloEarPea14}, $R_s = 10 h^{-1}
Mpc, \lambda_{\rm th} = 0.261$, the trend reverses,
$V \sim 2 V_{\rm  halo \; Lag}, \delta \sim 0.5 \delta_{\rm extrap}$.}

The change in these quantities using a different choice of
$\lambda^0_i$ can be seen in
Fig.~\ref{fig:bbks}, where we show mean
halo, filament and sheet maps for the same $R_s$ and $\lambda_{\rm
  th}$ values, but choice 3 for shear at the origin,
$\lambda^0_i = \bar{\lambda}_{i,{\rm
    extr}}$.
They are similar to their counterparts in Fig.~\ref{fig:aep}, but
slightly smaller, as is expected
with their smaller values of
$\lambda^0_i$ (see Fig~\ref{fig:lamcens}).
Choice 3 of shear at the origin also fixes $\partial \Phi=0$ at the
origin, implying values for the mean
Zel'dovich displacement for points in these structures.
In Fig.~\ref{fig:bbks}, black lines along each axis connect the
position of the initial edges, after they are shifted by their mean
Zel'dovich displacements ($\nabla \Phi$).  (No line is shown if the
displacement crosses the origin.  A displacement crossing the origin means
that the estimates of the final edge values are well outside of the
range of their validity.)  Under these displacements, the mean halo
contracts and the direction with the largest shear at early times
becomes the most contracted at late times, as expected and seen as
well in \citet{DesTorShe13}, see also \citet{LudBorPor14,Rob15}.
In addition, the mean filament and sheet both contract (only slightly
in some directions), while the void, not shown, expands in all
directions.
As mentioned earlier, the final time estimate for the linear
displacement of the initial region edges provides one (very rough)
estimate of the mean region's final size in the Lagrangian picture.
\footnote{A final overdensity
estimate can also be made from the ratio of the final to initial
volume in this picture.  This does not always work well in our
examples as the edges of structures be decreased beyond their
original distance from the
origin (implying the approximation has broken down long before
this point).}  

\begin{figure}
\begin{center}
\resizebox{3.in}{!}{\includegraphics{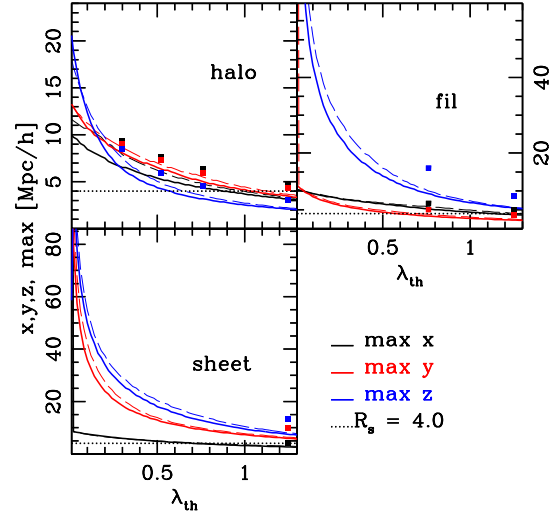}}
\end{center}
\caption{The extent of halos, filaments and sheets,
along the $x,y,z$ axes when 
$\lambda^0_i= \bar{\lambda}_{i,CA}$ (heavy solid lines)
and $\lambda^0_i= \bar{\lambda}_{i,{\rm extr}}$ (light dashed lines),
as a function of $\lambda_{\rm th}$.
Filled squares are
counterparts for halos,
filaments and sheets, at the thresholds shown, when
 $\bar{\lambda}_i =\bar{\lambda}_{i,CA}$, case 1. 
The smoothing scale $R_s = 4
h^{-1} Mpc$.  The largest eigenvalue is along $x$.
Examples of objects for case 1 and case 3 are shown in 
Fig.~\ref{fig:aep} and Fig.~\ref{fig:bbks} respectively.  
}
\label{fig:varlam}  
\end{figure}

More generally, the shapes of the mean objects in these
maps as the threshold $\lambda_{\rm th}$ changes can be seen in 
Fig.~\ref{fig:varlam}, for $R_s = 4 h^{-1} Mpc$ and our 3 choices of shear at the origin.
(Note that as these are the maximum of each coordinate that the actual
filament length is, for instance twice the maximum $z$ value.)
Solid points give the extent along the 3 axes for values of
$\lambda_{\rm th}$ where we solved choice 1
($\bar{\lambda}_i =
\bar{\lambda}_{i,CA}$), solid lines denote choice 2, mean maps
centered on $\lambda^0_i \bar{\lambda}_{i,CA}$, and the lighter dashed
lines are for choice 3, 
$\lambda^0_i = \bar{\lambda}_{i,{\rm extr}}$.
We only sampled central values for condition 1 because it required solving an implicit equation
for each $\lambda_{\rm th}$ and each different type of structure, and
again, we could not always find a solution. 

For a fixed smoothing scale (here $R_s$), raising the threshold $\lambda_{\rm th}$
in general raises the average
shears within objects, $\bar{\lambda}_i$.  As
can be seen, this simultaneously decreases the volume of
the associated mean region for any object beside a void.
The shapes of halos, i.e. ratios of extent along different axes, 
change strongly as the threshold $\lambda_{\rm th}$ varies.\footnote{
For other smoothings $R_s$, not shown here, we found that the sizes of
the structures along the different axes (in units of $R_s$ and
relative to those seen in Fig.~\ref{fig:varlam}) 
increase for lower $R_s< 4 h^{-1}Mpc$ and decrease for larger $R_s>4
h^{-1} Mpc$.  For example,
at low $\lambda_{\rm th}$, the size along a given coordinate
axis ($x,y,z$) for a given $\lambda_{\rm th}$ 
is
about twice as large (2/3 as large), in units of $R_s$, for
$R_s$ = 0.5 (19.0) $h^{-1} Mpc$.  The ratio of largest sheet edge to largest
filament edge is close to constant as $R_s$ changes, for $\lambda_{\rm
  th}>0$, but the filament/halo largest edge ratio changes (drops)
with
increasing $R_s$ until $\lambda_{\rm th} \sim > 0.2$.}

For choice 3 of shear at the origin, the trends of the mean Zel'dovich
displacements of the edges of the mean objects could also be studied.
We found that
the halo edges contract for all $\lambda_{\rm
  th}$ and that as $\lambda_{\rm th}$ increases the displacements go
through the origin and become negative, indicating the approximation
is well beyond its range of validity. 
For filaments and sheets, although 2 or 1 of the edges are/is displaced
quickly towards the origin even for small $\lambda_{\rm th}$, for the
other edge or edges, as $\lambda_{\rm th}$ increases, 
the displacement either is small at first (although
causing the displaced edge to move towards the origin), or causes the object to
first expand (the two larger sheet axes) before again having the
displaced edge move towards the origin.
The mean Zel'dovich displaced position of the edges
divided by the original position of the edges
takes roughly the same form for all choices of smoothing we studied.

\begin{figure*}
\begin{center}
\resizebox{3.in}{!}{\includegraphics{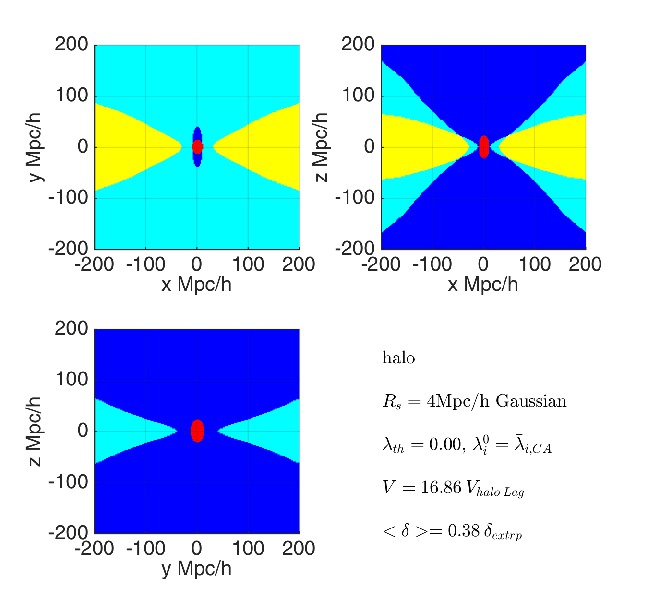}}
\resizebox{3.in}{!}{\includegraphics{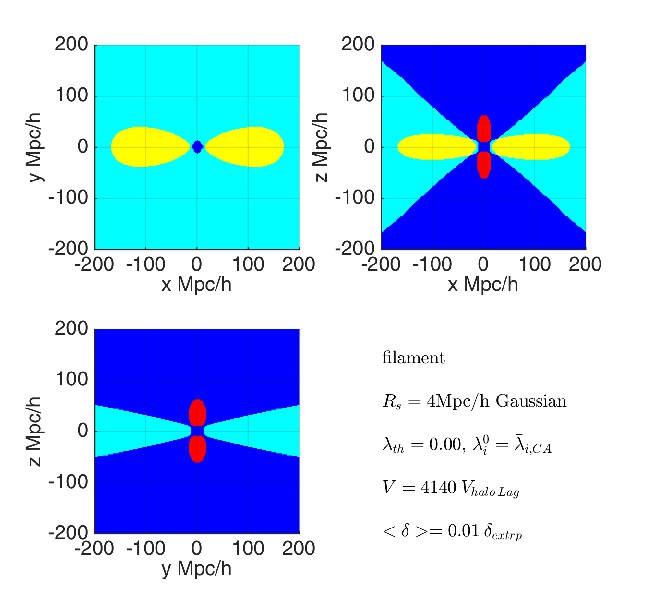}}
\resizebox{3.in}{!}{\includegraphics{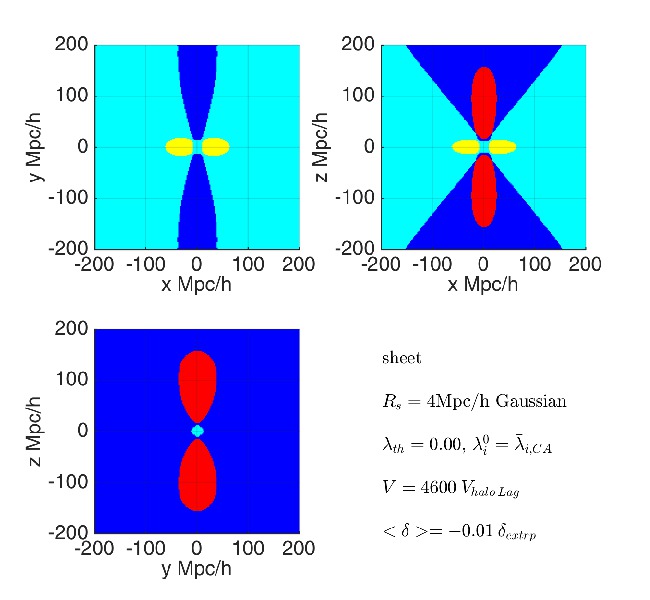}}
\resizebox{3.in}{!}{\includegraphics{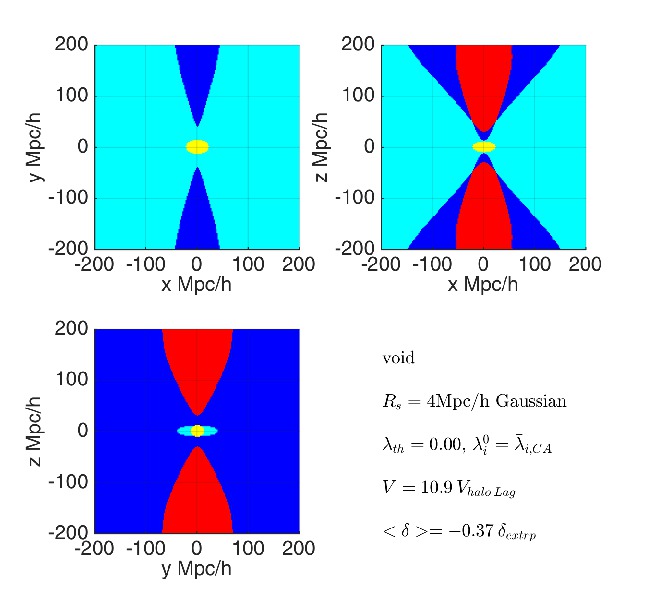}}
\end{center}
\caption{The extreme behavior for $\lambda_{\rm th} = 10^{-4}$ of mean regions around
$\lambda^0_i = \bar{\lambda}_{i,CA}$ (case 2) for a halo (top left)
filament (top right), sheet (lower left) and void (lower right) for smoothing
$R_s =4 h^{-1} Mpc$.  
Colors are as in Fig. ~\ref{fig:aep} and Fig.~\ref{fig:bbks}.  The 
larger axes scales are used in order to capture the increased volume of non-void
structures.
The halos in filament edges and
filaments in sheet edges visible for this $\lambda_{\rm th}$ decrease
as $\lambda_{\rm th}$ increases, eventually disappearing for
$\lambda_{\rm th} \sim 0.15$.   
 }
\label{fig:lam0}  
\end{figure*}

For many values of 
$\lambda_{\rm th}$ and $R_s$, the mean objects around average
points resemble
those in Figs.~\ref{fig:aep},\ref{fig:bbks} above, albeit with different
side lengths as just noted.
However, as $\lambda_{\rm th}$ tends to zero, not only do central
structures get much larger as seen in Fig.~\ref{fig:varlam}, but also
qualitatively different behavior appears.  Cross sections for the case
$\lambda_{\rm th} = 10^{-4}$ and case 2, $\lambda^0_i =
\bar{\lambda}_{i,CA}$, are shown in
Fig.~\ref{fig:lam0}.   The mean maps are shown for a much larger range
of scales, because
unlike the two earlier cases 
 shown in
Figs.~\ref{fig:aep},\ref{fig:bbks}, where
outside the central regions 
the shear quickly tends to a void, for this low
$\lambda_{\rm th}$ non-void
structures persist out to large distances.  (We consider
$\lambda_{\rm th} =10^{-4}$ rather than exactly zero as the functions
used to calculate the shears have limited accuracy.)
For this threshold, in part because voids and halos
only comprise 8\% of the total volume each, with sheets and filaments
taking 42\% each, the structures which are not the mean background
(not voids) are
extremely large.    In addition, unlike Fig.~\ref{fig:aep},
mean filaments have halos within their endpoints,
mean sheets have filaments within their outskirts, mean voids have sheets
outside of them, and so on, as can be seen.  
(The presence of halos
within the ends of filaments is characteristic of individual filaments
in simulations,
which are often found by looking for bridges between halos, e.g. \citet{Col07}.)
Around mean halos, the filaments fan out and do not
end within the scales ($\sim 800 h^{-1} Mpc$) we consider in 2 directions.
Increasing $\lambda_{\rm th}>0$, these extra embedded structures 
decrease in size and eventually disappear,
smoothly transforming into objects similar to
those shown in Fig.~\ref{fig:aep}.   For
$R_s = 4 h^{-1} Mpc$ this
transition happens around $\lambda_{\rm th} \sim 0.15$, for $R_s = 10 h^{-1}
Mpc$ this happens at an only slightly larger $\lambda_{\rm th} \sim 0.17$.
(As might be expected, this rich structure complicates using choice 1
to calculate
$\lambda^0_i$,
and is discussed
in Appendix \S\ref{sec:subtle}.)   The scatter around this mean configuration
is significant as well, as discussed in \S\ref{sec:scatter} below.

\subsection{Mean profiles along axes and scatter}
\label{sec:scatter}
Beyond shape, volume, extrapolated overdensity and mean Zel'dovich 
displacement of 
edges of each type of object, one can look at more details of the
mean shear itself, for instance the shear values along any given axis. 
For the mean objects in Figs.~\ref{fig:aep},\ref{fig:bbks}, values
along 3 axes for the 3 shear eigenvalues (and
the linearly extrapolated overdensity 
$\delta = \sigma \sum_i \lambda_i$) are shown in
Fig.~\ref{fig:profiles}.  
The dotted straight black line is the threshold $\lambda_{\rm th}$,
crossing this means changing structure classificiation 
(e.g. from a halo to a filament).
 The profiles for voids with $\lambda^0_i$
corresponding to central shear choices 2 and 3
($\lambda_{\rm th}=0.762$) are also shown, as these are explicitly calculable.
Central shear choices 1, 2 and 3 correspond to line types dot-dashed,
solid and dashed respectively.  The different choices of shear at the origin give similar
profiles for halos, but the filaments, sheets and voids show some 
large differences.  The void profile, when calculable (i.e. choices 2
and 3 at the origin), has
mass compensation at a distance away from the origin in the linearly
extrapolated over density as expected\footnote{We thank J. Peacock
  for suggesting we look for this.}.
Note that the mean shears take a very long time to approach their eventual
value of zero even though they fall below the threshold $\lambda_{\rm
  th}$ fairly quickly.  

\begin{figure}
\begin{center}
\resizebox{3.in}{!}{\includegraphics{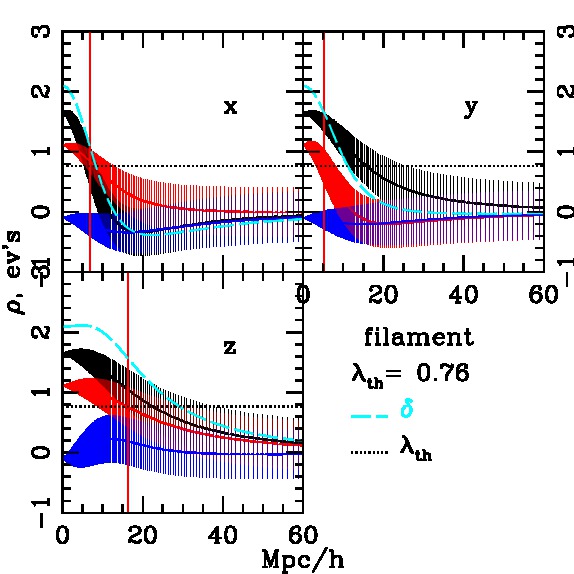}}
\resizebox{3.in}{!}{\includegraphics{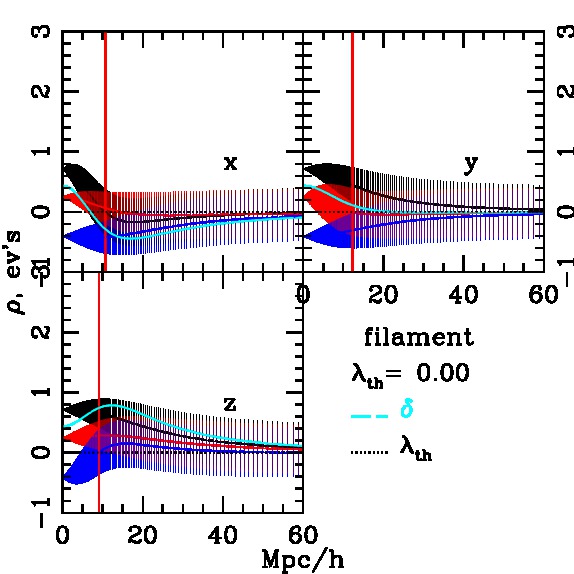}}
\end{center}
\caption{
Example of scatter around the mean profiles for a filament with
two different central values,
$\lambda^0_i$ set using criterion 1, and $\lambda_{\rm th} =
0.762$ at top,
using criterion 2 with
$\lambda_{\rm  th}=10^{-4}$
at bottom.   Cross sections for their mean maps are shown in 
Figs.~\ref{fig:aep},\ref{fig:lam0} respectively. 
As a function of distance from the origin, as 
the smoothing $R_s=4 h^{-1} Mpc$ for both, the
scatter is the same for both examples.   However, the effect on the
classification of the regions differs because of the difference in the
central values due to the difference in threshold $\lambda_{\rm th}$. 
The solid lines are the shear eigenvalue profiles $ \lambda_i (r)$
  (black, blue, red), and the filled regions show one sigma scatter along
 $x$, $y$ and $z$ axes.  The top filament profile
corresponds to the cross sections in Fig.~\ref{fig:aep} and
the profile in Fig.~\ref{fig:profiles}, the lower to
the filament cross section shown in
Fig.~\ref{fig:lam0}.
The dotted line again is the threshold $\lambda_{\rm th}$.  The
vertical solid red line is where the mean central filament transitions
along that axis to a sheet.
For low $\lambda_{\rm th}$, the scatter mixes different structures
even at large distances from the origin.
}
\label{fig:scatprof}  
\end{figure}

One other scale in the system is the scatter, the variance of the
shear around the mean values (calculated in Appendix
\S\ref{sec:shearcalc}, Eq.~\ref{eq:scatterexp}).\footnote{We
  thank V. Desjacques for suggesting we include this.}  
The variance depends only on $R_s$, the window function, and the
position relative to the constrained origin, independent 
of the central value of the mean configuration
and the threshold $\lambda_{\rm th}$.  It is simplest to interpret
along the 3 axes, where it starts at zero at the origin
 (where the shear constraint is imposed) and then
increases to a constant $\sim$0.2 at a distance $\sim 2 R_s +1$ for one
eigenvalue
and at a larger distance, $\sim 4.4 R_s + 5.5$, for the other two. 
(Calling the matrix in Eq.~\ref{eq:scatterexp} $C_{ij}$, where $i,j$
are each pairs of indices, the scatter shown is $1/\sqrt{
  (C^{-1})_{ii}}$ along $i$=11, 22 or 33 respectively.)
The effect of this scatter on the classification of a given region
depends upon the configuration at the origin and $\lambda_{\rm th}$.
We show in Fig.~\ref{fig:scatprof} 
the shear eigenvalue profiles along each of the 3 axes both
for the filament case seen in in Fig.~\ref{fig:profiles}), and for the
filament configuration with $\lambda_{\rm th} \sim 0$, but now also
show the envelope of the variance.  

In both examples, far from the origin 
the mean value of shear approaches zero, while the
scatter approaches its asymptotic value.  At 
large distances, if $\lambda_{\rm th}$ is high,
the
configurations close to the mean do not scatter enough to change their
classification from a void to anything else.  As $\lambda_{\rm th}$ drops,
a one-sigma fluctuation will mix many different kinds of structures 
even at large distances from the origin.  Thus, for instance,in
Fig.~\ref{fig:lam0} of the cross sections for low $\lambda_{\rm th}$,
beyond about 20 $h^{-1} Mpc$, the mean
configurations are within one-sigma of all of the structures (e.g., the
mean halo is within one-sigma of the mean filament, sheet and void).


Thus the different ways of constructing the mean maps, via different
constraints $C_0$ at the origin, provide, for the same $\lambda_{\rm th}$, 
similar objects in size and shape.
The mean central objects increase in size and change shape as the
threshold $\lambda_{\rm th}$ is decreased, becoming very complicated as the threshold
goes to zero. 
For very small $\lambda_{\rm th}$, the scatter strongly mixes the
different structures even far from the origin. The scatter about the mean configurations 
is independent of the central
value and approaches its asymptotic limit at larger distances from the
origin as $R_s$ increases.

\section{Collapsed Halo Interpretation}
\label{sec:collapse}
In general,  any smoothing $R_s$, and any threshold
$\lambda_{\rm th}$ can be used to define halos, filaments, sheets and
voids.  One can also ask how objects
in the smoothed map, such as halos, are related to other ways of
defining such objects (e.g. via gravitational collapse at later
times).   

As halos are the most well studied collapsed objects,
we investigated correspondences between our mean halos defined
using $\bar{\lambda}_i = \bar{\lambda}_{i,CA}$
and final collapsed halos such as are identified in numerical dark
matter simulations.
Initial conditions for halos and their final properties have been
extensively studied in
\citet{DalWhiBonShi08,Rob09,LudPor11,EliLudPor12,DesTorShe13,BorLudPor14,LudBorPor14}.
Final collapsed halos can be characterized by their mass and
the measured linearly extrapolated overdensity of their initial
regions.
We checked whether we could find, for each $R_s$, a $\lambda_{\rm th}$ 
such that our mean halo map overdensities matched
 $\delta_{\rm extrap}(R_s)$ from \citet{DesTorShe13} and 
$V_{\rm halo \; Lag} =M(R_s)/\rho_b$.  \footnote{The quantities being compared
are only approximate, as higher mass halos are
also included, as higher peaks, with smoothing $R_s$, 
and there are different mass definitions for
different quantities, with large conversion factors \citep{Whi01}.
For the former, the
average mass of all halos with $M \geq M(R_s) = M_+(R_s)$
and their average $\delta_+(R_s) = 1.686(1+0.2 \sigma_+(R_s)$
from integrating over the mass function, e.g., \citet{Tin08}, gives 
 $M_+(R_s)/M(R_s)>1.2$ until $R_s$ increases to 
$R_s \sim 9 h^{-1} Mpc$.
However $\delta_+(R_s)/\delta(R_s)
\sim 1$ for the smoothings $R_s=0.5-19.0$ which 
we considered.}  
We considered all three origin choices in \S\ref{sec:choices}.
They gave similar results: for smaller $R_s$, fixing $\langle \delta
\rangle$ to the value matching $R_s$ gave
a $\lambda_{\rm th}$ 
driven close to zero, or even below, and the Lagrangian volumes were much larger than
that associated with $M(R_s)$, and for larger $R_s$ the opposite happened.
For $R_s =4 h^{-1} Mpc$, for cases 1,2,3 respectively we found that
$\lambda_{\rm th} = (0.37,0.48, 0.43)$ and
$V = (2.1, 0.57, 1.0) M(R_s)/\rho_b$. (This can be compared to 
the average mass for $R_s$ and above at
this scale, $M_+(R_s) \sim
1.5 M(R_s)$.)  Given the approximations involved and the difficulty in
taking into account the higher mass halos, this might be a close
enough
agreement.  
In summary, only for one smoothing scale $R_s$ did it seem possible to
find a $\lambda_{\rm th}$ where the
mean halo had its volume close to
the Lagrangian volume $M(R_s)/\rho_b$,
for our 3 choices of mean object centers.

Initial condition halos do not always give final halos, and not all
halos at final times are tagged as halos in the initial conditions,
for
reasons which have been studied in detail in e.g.
\citet{DalWhiBonShi08,Rob09,LudPor11,EliLudPor12,DesTorShe13,BorLudPor14,LudBorPor14}.
For instance, external tidal fields might
destroy a initial local peak, or a peak might form at a
later time due to a merger.

If we take this choice of threshold as a preferred one for smoothing
$R_s = 4 h^{-1} Mpc$, because it gives a better connection to
collapsed halos (of one fixed mass) at late times, then we also can
consider the characteristic sizes for other objects for this
smoothing, from Fig.~\ref{fig:varlam}.   
However, this lack of direct connection to
collapsed halos (of one fixed mass) is not necessarily surprising.  
The mean objects we find for each kind of structure should be more
characteristic of the smoothed map than, e.g., for halos, the
collapsed objects which can form after full nonlinear evolution.

For modified constraints on 
both $\lambda_{\rm th}$ and $\delta$
and taking $\lambda^0_i = \bar{\lambda}_{i,CA}$,
\citet{PapShe13} compared the analytically predicted 
density anisotropy at large, i.e. close to linear,
distances to that produced by 
a stack of numerically simulated halos, finding rough agreement.
Properties of the 
mean objects for the origin constraints considered by
\citet{PapShe13} are given in the Appendix \S\ref{sec:papsheex}.

\section{Summary and Discussion}
\label{sec:conclude}

A spatial distribution of Gaussian random fields can be classified into
halos, filaments, sheets and voids using a smoothing scale $R_s$
and threshold $\lambda_{\rm th}$.  These can be associated with
linearly evolved initial conditions or with the initial conditions
themselves for structure formation.   Many smoothing scales and
thresholds have been chosen for use in different contexts.
The comparison of fractions of volume and mass in the different
structures has shown some agreement between analytic calculations of
initial Gaussian random fields (specifically their linearly extrapolated
values at the present time)  and numerical simulations of
the full nonlinear evolution (e.g., \citet{AloEarPea14}).

We explored further properties of cosmic structures in the Gaussian
random field configurations, for the same classification and
parameters used by \citet{AloEarPea14}.
We used the properties of Gaussian random fields, combined with the
power spectrum, to calculate
properties of mean halos, filaments, sheets and voids
 with three choices of central shears $\lambda^0_i$.  These
central shears are related either to
 the mean shears for each type of object, or to the volume average of
 the extremal shear values for each object.
 For low $\lambda_{\rm th}$ the
mean structures are very complicated, while at larger $\lambda_{\rm th}$
the mean
structures centered at the origin smoothly transition to the
mean background through the sequence halo to filament to sheet to
void.
The mean halo has approximately similar extent in all three
directions, the mean filament has one larger direction and the mean sheet is
small along one direction and larger along the other two.
We showed examples of mean structures at large and small 
$\lambda_{\rm th}$ and discussed
trends of their shapes with changes in threshold and in smoothing.
One method of choosing the central shear was an implicit condition,
which we could not always solve.
When the central point is taken to be an extremal point ($\lambda_i^0
= \bar{\lambda}_{i,{\rm extr}}$), there is a natural value for the
Zel'dovich displacement at the origin and thus one can also calculate
the mean Zel'dovich displacement of the edge points of the mean objects,
giving a
final linear approximation to the size of each of these objects.  
For some $\lambda_{\rm th}$, the estimated Zel'dovich change in the mean
filaments and sheets is small, while the mean filament and sheets are
large relative to the smoothing scale (and sometimes the nonlinear scale),
suggesting that these characteristic sizes might persist in the
nonlinear maps in some form as well.

We also
characterized the scatter around the mean configurations, which
tends to a constant at a small multiple of $R_s$, the smoothing.  The
effect of scatter 
on the classification of structures around the mean background
depends in part on $\lambda_{\rm th}$, we showed as an
example the scatter for
mean filaments for two values of $\lambda_{\rm th}$.

There is not a clear identification of the mean halos with collapsed
halos or collapsed halos above some mass. 
Matching the expected average linearly extrapolated overdensity
for a given smoothing $R_s$ implies a threshold $\lambda_{\rm th}$,
and leads to a Lagrangian volume (final estimated mass)
which is too small (large) relative to $M(R_s)$ for large (small) $R_s$, although there might be a sweet spot 
near $R_s \sim 4 h^{-1} Mpc$.  These estimates were only approximate
as different mass definitions are used in the quantities being compared.

As these mean objects are an additional characterization of
initial Gaussian random field configurations and smoothed
maps, it would be very interesting to compare them to
mean objects in simulations with the same smoothing and threshold,
both at initial and at late (fully evolved) times.  
 This comparison is nontrivial.  
Although the division of halo, filament, sheet and
void regions is direct in simulations, as are their volume and
mass fractions, the separation 
 into specific distinct objects, e.g. different filaments, is not immediate.
 Splitting halo,
filament, sheet and void regions into objects so that their stack gives our mean
objects
might be a natural way of defining such individual objects in the full
map, but presumably additional assumptions will be required.  The
stacking might be most naturally associated with doing a volume
average over extremal points as in our choice 3 of origin shear.

This work is in a different direction than the
extremely interesting research on finding the probability distribution
of initial conditions which are consistent with a given final measured nonlinear
configuration (e.g. the line of study pursued in \citet{JasLecWan15,LecJasWan15a,LecJasWan15b,LecJasWan15c}) although both kinds
of studies involve cosmic web classifications of initial conditions
and their inferred counterparts at late times.  

Several free parameters appeared in the construction, including
the threshold $\lambda_{\rm th}$, the smoothing $R_s$ and
the central values $\lambda^0_i$.  We chose a few different
$\lambda^0_i$, however others are possible.
Connections to other natural scales in the initial conditions 
would be interesting to
further pursue as well, such as those discussed in \citet{Sha09} and
\citet{Pog09}.

JDC thanks J. Peacock, A. Pontzen and J. Whitehouse for suggestions,
and especially M. White 
for numerous discussions.  We thank N. Dalal, V. Desjacques,
F. Leclercq, M. Neyrinck
and the anonymous referee 
for helpful comments on
the draft.

\appendix
\section{Shear calculation details}
\label{sec:shearcalc}
The mean shear $\langle \xi_{ij}({\bf r})
|C_0\rangle$
surrounding the origin with fixed boundary conditions 
$\langle \xi_{ij}|C_0\rangle  = \lambda^0_i \delta_{ij}$ can
be calculated as described in \citet{PapShe13}  (based
in part upon \citet{BBKS}).  Their focus was on 
$\langle \delta ({\bf r}) |C_0\rangle$, for only one of the boundary
conditions $C_0$ which we consider, so we give more details for
completeness.  A more general angle-averaged probability 
distribution $P(\vec{\xi}(r_1),\vec{\xi} (r_2),r )$ 
was found by \citet{DesSmi08}.

Defining $\vec{\xi}({\bf r}) =(\xi_{11}({\bf r}),
\xi_{22}({\bf r}),\xi_{33}({\bf r}),
\xi_{12}({\bf r}),\xi_{23}({\bf r})$,\\$\xi_{13}({\bf r}))$, 
and $\vec{\xi} \equiv \vec{\xi}(0)$,
\citet{PapShe13} use
the Gaussianity of
$p(\vec{\xi}(r)|\vec{\xi})$ to get
\begin{equation}
\langle \vec{\xi}({\bf r})|\vec{\xi} \rangle = \langle \vec{\xi}({\bf r}) \vec{\xi} \rangle \langle
\vec{\xi} \otimes
\vec{\xi}\rangle^{-1} \vec{\xi} \; .
\label{eq:shortexp}
\end{equation}
which for conditions $C_0$ at the origin then give
\begin{equation}
\langle \vec{\xi}({\bf r})|C_0\rangle = \langle \vec{\xi}({\bf r}) \vec{\xi} \rangle \langle
\vec{\xi} \otimes
\vec{\xi}\rangle^{-1} \langle \vec{\xi} | C_0 \rangle \; .
\label{eq:fullexp}
\end{equation}
The Zel'dovich displacement mean value, shown for central shear choice
3 in Fig.~\ref{fig:bbks} can also be calculated
analogously to that for $\vec{\xi}({\bf r})$ when the displacement at
the origin is fixed to be zero (as it is for choice 3 of central shear):
\begin{equation}
\langle \Psi^Z_i({\bf r})|C_0 \rangle = \langle \Psi^Z_i({\bf r}) \vec{\xi} \rangle \langle
\vec{\xi} \otimes
\vec{\xi}\rangle^{-1} \langle \vec{\xi} | C_0 \rangle \; .
\label{eq:psiexp}
\end{equation}
The variance of $\vec{\xi}({\bf r})-\langle \vec{\xi}({\bf
  r})|C_0\rangle$
is \citep{BBKS}:
\begin{equation}
\langle \vec{\xi} \otimes
\vec{\xi}\rangle- \langle \vec{\xi}({\bf r}) \vec{\xi} \rangle \langle
\vec{\xi} \otimes
\vec{\xi}\rangle^{-1} \langle \vec{\xi} | \vec{\xi}({\bf r}) \rangle \; .
\label{eq:scatterexp}
\end{equation}
which is zero at the origin (where there is a fixed constraint) and
then increases, asymptotically reaching $\langle \vec{\xi} \otimes
\vec{\xi}\rangle$.

These give the mean value of the shear,  
Zel'dovich displacement when $\partial_i
\Phi=0$ at the origin, and scatter at position ${\bf r}$ when
the shear at the origin is set by the boundary condition
$C_0$, which we explore in the body of this paper.   The three central
shear
values $\lambda^0_i$ which we consider are in \S\ref{sec:choices}.

To evaluate Eq.~\ref{eq:fullexp}, Eq.~\ref{eq:psiexp} and Eq.~\ref{eq:scatterexp},
we follow \citet{PapShe13} in
choosing the shear at the origin to be diagonal $\vec{\xi}(0) \equiv
\vec{\xi} = \vec{\xi}_D =(\xi_{11},\xi_{22},\xi_{33})$, and 
substitute $\langle \xi_D \otimes
\xi_D\rangle^{-1}$:
\begin{equation}
\begin{array}{l}
\langle  \xi_{ij}({\bf r})|C_0 \rangle = \\ 
(6 \langle \xi_{ij}({\bf r}) \xi_{11} \rangle -3/2 \langle \xi_{ij}({\bf r})
\xi_{22} \rangle -3/2 \langle \xi_{ij}({\bf r}) \xi_{33} \rangle)\langle
\xi_{11}|C_0\rangle \\
+(-3/2 \langle \xi_{ij}({\bf r}) \xi_{11} \rangle +6 \langle \xi_{ij}({\bf r})
\xi_{22} \rangle -3/2 \langle \xi_{ij}({\bf r}) \xi_{33} \rangle)\langle
\xi_{22}|C_0\rangle \\
+(-3/2 \langle \xi_{ij}({\bf r}) \xi_{11} \rangle -3/2 \langle \xi_{ij}({\bf r})
\xi_{22} \rangle +6\langle \xi_{ij}({\bf r}) \xi_{33} \rangle)\langle
\xi_{33}|C_0\rangle \; . \\ 
\end{array}
\label{eq:inxic}
\end{equation}
The expression for $\Psi^Z_i$ can be found by replacing
$\xi_{ij}({\bf r})$ by $\Psi^Z_i({\bf r})$.

For the scatter, Eq.~\ref{eq:scatterexp},
\begin{equation}
\begin{array}{l}
\langle  (\xi_{ij}({\bf r})-\langle  \xi_{ij}({\bf r})|C_0 \rangle) (\xi_{kl}({\bf r})-\langle  \xi_{kl}({\bf r})|C_0 \rangle)
\rangle
 = \\ 
(6 \langle \xi_{ij}({\bf r}) \xi_{11} \rangle -3/2 \langle \xi_{ij}({\bf r})
\xi_{22} \rangle -3/2 \langle \xi_{ij}({\bf r}) \xi_{33} \rangle)\langle
\xi_{11}| \xi_{kl}({\bf r})\rangle \\
+(-3/2 \langle \xi_{ij}({\bf r}) \xi_{11} \rangle +6 \langle \xi_{ij}({\bf r})
\xi_{22} \rangle -3/2 \langle \xi_{ij}({\bf r}) \xi_{33} \rangle)\langle
\xi_{22}|\xi_{kl}({\bf r}) \rangle \\
+(-3/2 \langle \xi_{ij}({\bf r}) \xi_{11} \rangle -3/2 \langle \xi_{ij}({\bf r})
\xi_{22} \rangle +6\langle \xi_{ij}({\bf r}) \xi_{33} \rangle)\langle
\xi_{33}|\xi_{kl}({\bf r})\rangle \\
+ 15 (\langle \xi_{ij}({\bf r}) \xi_{12} \rangle \langle \xi_{12}
\xi_{kl}({\bf r})  \rangle \\
+\langle \xi_{ij}({\bf r}) \xi_{13} \rangle \langle \xi_{13}
\xi_{kl}({\bf r})  \rangle +
\langle \xi_{ij}({\bf r}) \xi_{23} \rangle \langle \xi_{23} \xi_{kl}({\bf r})  \rangle )
\end{array}
\label{eq:scatexplicit}
\end{equation}  
This is independent of the value at the origin, and the threshold
$\lambda_{\rm th}$.

The correlations 
$\langle \xi_{ij}({\bf r}) \xi_{mm} \rangle$ are
\begin{equation}
\begin{array}{ll}
\langle \xi_{ij}({\bf r}) \xi_{mm}\rangle&= \Psi_1(r) \hat{r}_i \hat{r}_j
\hat{r}_m^2 \\ 
&+\Psi_3(r)(2 \hat{r}_i \hat{r}_m \delta_{jm} 
+2\hat{r}_j \hat{r}_m \delta_{im} \\
&+ \hat{r}_i \hat{r}_j  +\hat{r}_m^2 
\delta_{ij}) \\
& + \Psi_5(r) (\delta_{ij} + 2
\delta_{im}\delta_{jm}) \; ,
\end{array}
\end{equation}
for $mm = 11,22,33$, and
\begin{equation}
\frac{1}{\sigma}\langle \Psi^Z_i ({\bf r}) \xi_{mn} \rangle =
\Psi_4 (r) \hat{r}_i \hat{r}_m \hat{r}_n
+ \Psi_2(r) (\hat{r}_i \delta_{mn} +\hat{r}_m \delta_{in} +\hat{r}_n
\delta_{im} ) \; .
\label{eq:psixi}
\end{equation}

\begin{figure}
\begin{center}
\resizebox{3.5in}{!}{\includegraphics{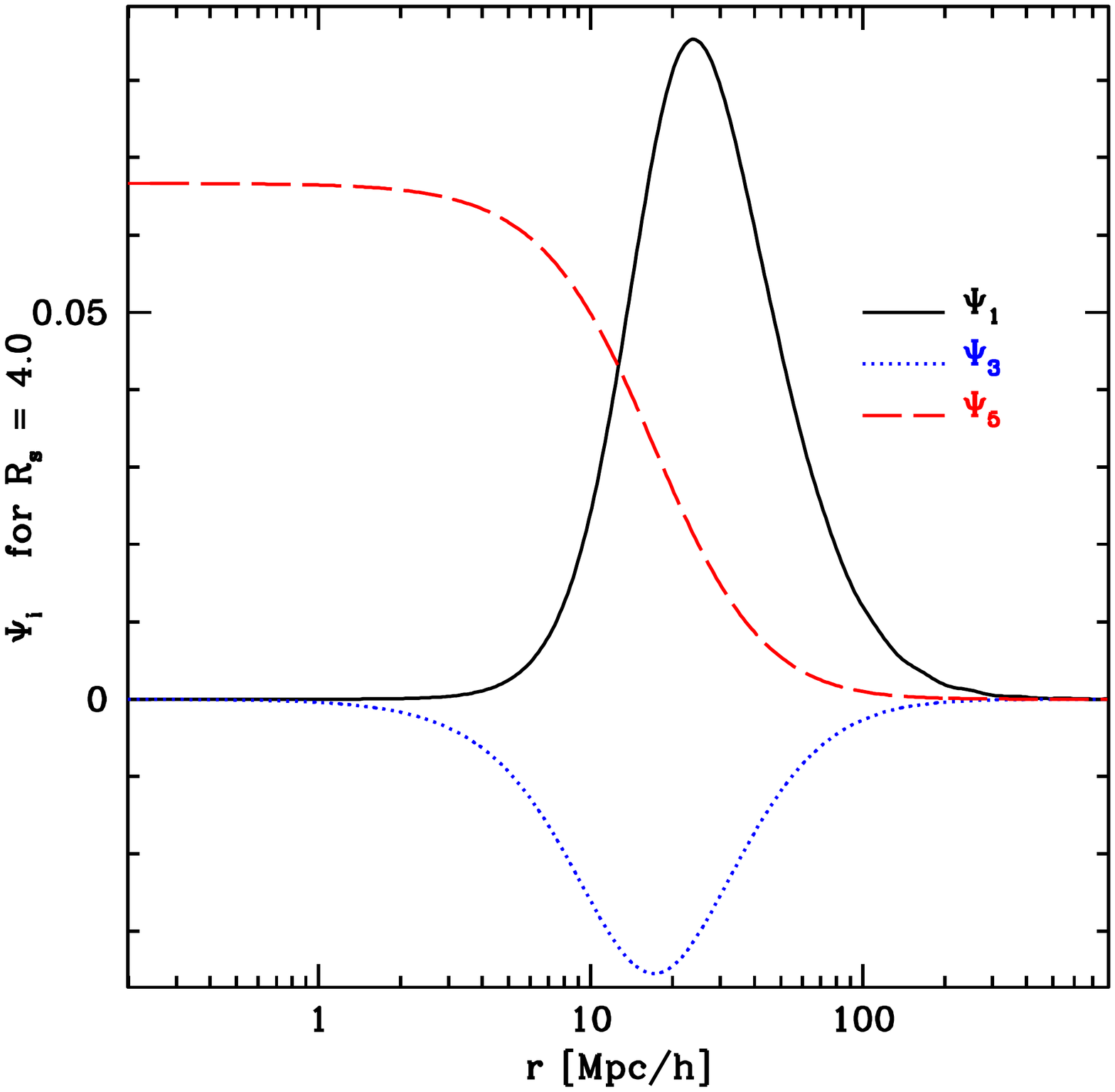}}
\end{center}
\caption{The three kernel functions $\Psi_1,\Psi_3,\Psi_5$
(Eq.~\ref{eq:threepsi}) used to calculate the mean shear
properties, for
a Gaussian window function with smoothing  $R_s = 4 h^{-1} Mpc$.
The natural scale is about 10-20 $h^{-1} Mpc$ comoving, larger than
the $R_s$ in part due to the softness of the cutoff in the
window function.
}
\label{fig:threepsis}  
\end{figure}

The basis functions $\Psi_i(r)$ depend on the primordial 
matter power spectrum  $\Delta_\delta^2(k) = k^3 P_L(k)/(2 \pi^2)$.
We use a power spectrum for
a $\Lambda$CDM universe with $(\Omega_m,\Omega_bh^2,h,\sigma_8,n) =
(0.31,0.022,0.68,0.83,0.96)$ \citep{Pla13}, calculated with the
\citet{EisHu99} transfer function.
For the $\Psi_i(r)$ associated with $\xi_{ij}$ (appendix of \citet{Des08}),
\begin{equation}
\begin{array}{ll}
\Psi_1(r)&= \frac{1}{\sigma^2}\int_0^\infty  d \ln k \Delta_\delta^2(k) W_R(k)^2j_4(kr) \\
\Psi_3(r)&= \frac{1}{\sigma^2}\int_0^\infty  d \ln k \Delta_\delta^2(k) W_R(k)^2[-\frac{1}{7} j_2(kr) -\frac{1}{7} j_4(kr)] \\
\Psi_5(r)&= \frac{1}{\sigma^2}\int_0^\infty  d \ln k
\Delta_\delta^2(k)W_R(k)^2 [\frac{1}{15}
j_0(kr) \\ &+ \frac{2}{21}j_2(kr) +\frac{1}{35} j_4(kr)] \; . \\
\end{array}
\label{eq:threepsi}
\end{equation}
These 
are shown in Fig.~\ref{fig:threepsis} for 
smoothing $R_s = 4
h^{-1} Mpc$, the case considered in most detail in this note.
There are features at scales a few times $R_s$.
For $\Psi^Z_i$'s correlations, the associated basis functions are
(e.g., \citet{Whi14}):
\begin{equation}
\begin{array}{lll}
\Psi_2(r) &=&\frac{1}{\sigma^2}\int_0^\infty  d \ln k
\Delta_\delta^2(k) \frac{W_R(k)^2}{k}(-\frac{1}{5}j_1(kr) -\frac{1}{5}j_3(kr))\\
\Psi_4(r) &=&\frac{1}{\sigma^2}\int_0^\infty  d \ln k
\Delta_\delta^2(k) \frac{W_R(k)^2}{k}j_3(kr) \; . \\
\end{array}
\end{equation}
These are used to calculate the mean displaced edges of the cosmological
structures shown as solid black lines in Fig.~\ref{fig:bbks}.

\section{Subtleties in fixing $\bar{\lambda}_i = \bar{\lambda}_{i,CA}$}
\label{sec:subtle}
Method 1 of assigning $\lambda^0_i$ (in \S\ref{sec:choices})
constrains its value implicitly by requiring the average shear in a
mean object of a certain kind, centered at the origin, to 
be equal to the mean value of shear
for that kind of object in Eq.~\ref{eq:dorav}.  In addition to not always being able
to solve this constraint by systematic trial and error, 
there are also conceptual subtleties
with this approach, although
there is an intuitive appeal to
requiring the average
of the shear within the central mean object in our construction to be
the analytically calculated average of the shear within that type of object, $\bar{\lambda}_{i,CA}$.
One complication was already noted for voids, 
as constructing a mean void volume is
challenging when there is no boundary between a mean ``void''
configuration and the mean background, i.e. no mean void ``size''.
In practice, restricting to halos, filaments and sheets, we took objects centered
on each kind of structure
and calculated the average shear within these objects for a given
$\lambda_{\rm th}$.  We then varied the central shear until the mean
shear within each object  was the one given by Eq.~\ref{eq:dorav} for
that object.

However, the mean maps often have more than just
the central type of object and the void structure in them; all four
structures
can be present in a mean map centered on any one type of structure.
So when one calculates the mean shear in the mean objects of a certain
structure, one
possible way is to include the contributions from all the different
maps where that structure appears.  That is all regions with that
given
structure, even if not part of the central object,
should be included as well.
If the average in the mean maps is chosen to include the contributions
from these other regions, the calculation of mean shear becomes more complicated.

We give an example by showing how adding these regions to calculate
the mean shear can affect the calculation of
mean shear for filaments.
Generally, a mean halo at the origin also has filament 
regions around it.
So in addition to the mean filament region around the origin, when the origin is a
filament, the mean of filament shears including other regions would
include the shear contribution from filament regions around the mean central halo,
calculated for the same threshold.
The relative contributions of these filaments in the different maps,
filaments around mean halos, around central mean filaments, (and
sheets and voids in some cases), can
be estimated if one breaks down the volume fraction for each kind of
object into contributions from each configuration: i.e. filaments when
there is a mean halo at
center, when there is a mean filament at center and when there is a
mean sheet at center.  This is
\begin{equation}
\begin{array}{ll}
V_{\rm tot}
f_{Vf} = &N_{\rm halo} \bar{V}_{\rm fil \in halo} + N_{\rm fil}\bar{V}_{\rm
  fil} +N_{\rm sheet} \bar{V}_{\rm fil \in sheet} \\
& +N_{\rm voids} \bar{V}_{\rm fil \in voids} \; .
\end{array}
\label{eq:filvol}
\end{equation}
Here the volume fractions
$f_{Vh},f_{Vf},f_{Vs},f_{Vv}$ of each structure are available
from Eq.~\ref{eq:dorav}, 
and $\bar{V}_{\rm fil \in halo}$ is the
filament volume associated with each mean halo, etc., found by
constructing the mean halo, etc. and measuring the volume.
These volumes depends upon the central shears which one is trying to
find.  This is especially challenging if
contributions to the mean filamentary shear are large from any mean
map besides that centered on the mean filament, and gets even more
self-referential once filaments appear in the outskirts of
sheets or halos appear in the outskirts of filaments, i.e. the last
two terms $\bar{V}_{\rm fil \in sheet} ,\bar{V}_{\rm fil
  \in voids} \ne 0$.

For some $\lambda_{\rm th}$ these additional terms
$N_{\rm halo} \bar{V}_{\rm fil \in halo} +N_{\rm sheet} \bar{V}_{\rm fil \in sheet} 
+N_{\rm voids} \bar{V}_{\rm fil \in voids}$ 
 seem small.  Then the
calculation ignoring these other terms, that is, averaging the filament shear only over the mean
object centered on the origin which is a filament, seems justified
even in this more complicated interpretation (and 
similarly for sheets and halos in their respective cases).
For filaments these other terms can be estimated for size by the
following method.
For $\lambda_{\rm th}$ large enough,
the mean halo structure only appears when it is the central object, so
$\lambda^0_i$ can be found for the mean halo by imposing $\lambda^0_i=
\bar{\lambda}_{i,CA}$ for a mean halo centered at the origin.
Also, again for $\lambda_{\rm th}$ large enough,
$\bar{V}_{\rm fil \in sheet} =\bar{V}_{\rm fil \in voids}=0$.  Then
filaments only appear around the central mean halo and as the central mean
filament.
So we have:
\begin{equation}
\begin{array}{ll}
V_{\rm tot}f_{Vh} & =N_{\rm halo} \bar{V}_{\rm halo} \\
V_{\rm tot}f_{Vf} &=N_{\rm halo} \bar{V}_{\rm
  fil \in halo}
+ N_{\rm fil} \bar{V}_{\rm fil} \; .
\end{array}
\end{equation}
and the filamentary volume around the mean halo,
$\bar{V}_{\rm
  fil \in halo}$ is determined by the mean halo, already in hand.  
As we also know $f_{Vh}/f_{Vf}$, we can constrain
\begin{equation}
\frac{N_{\rm halo} V_{\rm
  fil \in halo}}
{N_{\rm fil} V_{\rm fil} } =\frac{
V_{\rm
  fil \in halo }}
{\frac{f_{Vf}}{f_{Vh}} V_{\rm halo}- V_{\rm fil \in halo}}  \; .
\end{equation}

This can be used to see if the fraction of mean filament 
shear due to filaments around halos is large relative to that
due to a filament centered at the origin.
For $R_s = 4 h^{-1} Mpc$ halos and $\lambda_{\rm th} = 0.762$,
$V_{\rm fil \in halo } \sim
2 V_{\rm halo }$ and $f_{Vf}/f_{Vh} = 1/72$, giving a fraction of
filament volume due to
filaments
around halos of
about 1/35, which we thus neglected
in calculating the filament central value shown in
Fig.~\ref{fig:aep}. 
Repeating the argument for sheets, for this threshold and
smoothing, the relatively large 
sheet to filament volume fraction $f_{Vs}/f_{Vf}\sim 1/14$,
combined with the large $\bar{V}_{\rm sheet \in fil}$, makes
$\bar{V}_{\rm sheet \in fil}$ too large to reasonably neglect in
this interpretation.\footnote{We could also not find a
$\lambda^0_i$ (by trial and error) to meet the constraint
$\bar{\lambda}_i = \bar{\lambda}_{i,CA}$
 by considering the volume centered
on the sheet alone.  More generally, configurations where any
$\bar{\lambda}_{i,CA} <0$ were often quite hard to find via trial and
error.
This happens, for $R_s = 4 h^{-1} Mpc$, at $\lambda_{\rm th} < \sim 0.6$ for filaments and
$\lambda_{\rm th}< \sim 1.2$ for sheets.}
In Fig.~\ref{fig:aep} we thus considered instead $\lambda_{\rm th} = 1.25$ to
impose $\bar{\lambda}_i =\bar{\lambda}_{i,CA}$ in a sheet configuration.
(For the other scale studied
in detail by \citet{AloEarPea14}, $R_s = 10h^{-1} Mpc$,
filament contributions around halos and the filament to halo fraction (1/12)
were both too large for their $\lambda_{\rm th}=0.261$ to 
permit neglect of $\bar{V}_{\rm fil \in halo}$.) 
One could include these subleading contributions directly instead of
going to regimes where they can be neglected, as well, in some cases,
when halos are not found in filament edges, and filaments are not
found in sheet edges.  It would be
interesting to further explore this interpretation, although the
issues with voids might make it difficult to push it very far.

This mean halo/filament/sheet interpretation and 
argument also gives a characteristic mean halo, filament and sheet
volume 
and
an expression for $N_{\rm fil}$ in terms of
the number of halos, the volume fractions of filaments and halos and
the mean volumes of filaments and halos.  For the simplest case, for instance,
$N_{\rm fil} = \frac{N_{\rm halo}
}
{V_{\rm fil } } (
\frac{f_{Vf}}{f_{Vh}} V_{\rm halo}- V_{\rm fil \in halo})$, and
analogously for sheets.    

\begin{table*}
\centering
\begin{tabular}{|l|c|c|c|c|c|c|}
\hline
$M(R_s)/h^{-1}M_\odot$&$R_s/h^{-1}Mpc$&$\lambda_{\rm
  th}$&\shortstack{V $\rho_b$\\$h^{-1}M_\odot$}&
x, y, z max& $\langle \delta \rangle$
&\shortstack{ $\lambda^0_i$ \\ $\bar{\lambda}_i$} \\ \hline
halo 1.0e+14 &4.2 & 0.54 & 4.70e+13 & 5.52 6.67 4.36 & 2.03 &\shortstack{ 1.35,  0.93, 0.65 \\  1.22
  0.85 0.58} 
\\ \hline
halo 5.0e+14 &7.2 & 0.81 & 9.74e+13 & 5.93 7.00 5.65  & 1.79 &\shortstack{ 1.63,  1.20, 0.92 \\  1.53 1.14 0.86} 
\\ \hline
halo 1.0e+15 &9.0 & 0.98 & 1.03e+14 & 5.52 6.36 6.22  & 1.74 &\shortstack{ 1.84,  1.39, 1.09 \\  1.77 1.35 1.04} 
\\ \hline
void 1.0e+14 &4.2 & 0.00 & 1.20e+13 & 4.06 3.13 2.59  & -2.88 &\shortstack{ -0.77,  -1.30, -1.84 \\  -0.74 -1.26 -1.77} 
\\ \hline
void 5.0e+14 &7.2 & 0.00 & 1.52e+13 &3.97 3.36 2.98  & -2.83 &\shortstack{ -1.36,  -1.90, -2.42 \\
  -1.33 -1.87 -2.38} \\ \hline
void 1.0e+15 &9.0 & 0.00 & 1.69e+13 &3.97 3.52 3.13  & -2.82 &\shortstack{ -1.75,  -2.28, -2.81 \\  -1.72 -2.25 -2.77} 
\\
\hline
\end{tabular}
\caption{Properties for
halos (top) and voids (bottom)
for objects using the smoothing scales $R_s$ and thresholds
$\lambda_{\rm th},\delta$ of \citet{PapShe13}.  These objects have
both a $\lambda_{\rm th}$ and $\delta$ constraint at the origin.
For halos, $\lambda_{\rm th}= 0.41/\sigma(R_s)$, $\delta > 1.686/\sigma(R_s)$.
For voids, $\lambda_{\rm th} = 0$, $\delta  < -2.8/\sigma(R_s)$.
The mass of the mean halo volumes tends to be smaller than that
associated
with the smoothing scale.}
\label{tab:pshalos}
\end{table*} 

\section{Stacked halos and voids with a different $\lambda^0_i$ constraint}
\label{sec:papsheex}
This method for calculating mean shear was used by
\citet{PapShe13} to find the mean
shear due to a given configuration at the origin. They took the origin
to correspond to 1) the sum of all halos with mass above some smoothing
scale and 2) all voids associated with a smoothing scale,
at large distances and final times.
To find the central values of shear, they
imposed conditions both on
the eigenvalues of the shear individually and on their sum (the
density).  We give here properties of the 
mean configuration maps for their boundary
conditions at the origin.  Note that we consider the mean smoothed
values given their boundary conditions (that is we smooth the shear
both at the origin and at the position it is being measured).

They chose $\lambda^0_i =
\bar{\lambda}_{i,RS}$, where the latter was
found by integrating Eq.~\ref{eq:dorav} with
either their halo or void constraints but with an additional
constraint in the measure, a step function in the density.  For halos,
$\lambda_{\rm th} (\sigma) = 0.41/\sigma$ and 
 $\sum_i \lambda_i > 1.686/\sigma$, for voids,
$\lambda_{\rm th}(\sigma) =0 $ and
 $\sum_i \lambda_i < -2.8/\sigma$.  This is analogous to our choice
of central shear 2, except for the additional density constraint.
In addition, voids and halos have
different thresholds $\lambda_{\rm th}$.

For the comparison with simulations,
as $R_s$ increases, $\sigma(R_s)$ decreases and $\lambda_{\rm th}$ 
increases, so halos with a larger smoothing
(higher mass) will also obey the constraints for a lower $R_s$ (this
was used in this note in \S\ref{sec:collapse}).  Thus for halos,
\citet{PapShe13}
interpret the resulting shear at the origin $\lambda^0_i$ as due to the sum of
halos with mass associated with $R_s$ and above.  They take
simulated halos with masses corresponding to $R_s$ and above
in a numerical simulation, stack them, and compare measured and 
analytically predicted anisotropies
in density at large distances, finding reasonable agreement.

They consider masses $10^{14}, 5 \times 10^{14}, 10^{15}
h^{-1} M_\odot$, from which we find $R_s$.  
In Table 1, 
we show halo and void configurations they used,
with the corresponding $R_s, \lambda_{\rm th}$,
volume ($\sim M$), shape,
average $z=0$ (Eulerian) overdensity,
and
$\lambda^0_i$, $\bar{\lambda}_i$.
We use a Gaussian window function.

Interestingly, for these constraints, the central value of shear and
the average value in the halo or void volume are very close. 
The density cut is a stringent 
constraint on the shear and the average density in the regions is close to the
minimum (maximum for voids) imposed density.  
The associated volumes for the halos correspond to smaller masses
than the Lagrangian volumes of the halos associated with $R_s$.

\end{document}